\documentclass[prx,twocolumn,aps,epsf,showpacs,superscriptaddress,longbibliography]{revtex4-2}

\usepackage{amsmath}
\usepackage{amssymb}
\usepackage{graphicx}
\usepackage{dsfont}
\usepackage{subfigure}
\usepackage{hyperref}
\usepackage{float}
\usepackage{wasysym}
\usepackage{bbm}
\usepackage{comment}
\usepackage{pdfpages}
\usepackage[normalem]{ulem}
\makeatletter
\AtBeginDocument{\let\LS@rot\@undefined}
\makeatother

\usepackage{pgffor}

\newcommand{\ket}[1]{|#1 \rangle}
\newcommand{\bra}[1]{\langle#1|}

\newcommand{\up}{\uparrow}
\newcommand{\down}{\downarrow}
\renewcommand{\>}{\rangle}

\renewcommand{\[}{\left[}

\renewcommand{\vec}[1]{\boldsymbol{#1}}

\hypersetup{colorlinks,citecolor=blue,linkcolor=blue,urlcolor=blue}

\newcommand{\N}{\mathcal{N}}

\begin{document}

\title{A Race Track Trapped-Ion Quantum Processor}


\author{S. A. Moses}
\thanks{These authors contributed equally to this work.}
\author{C. H. Baldwin}
\thanks{These authors contributed equally to this work.}
\author{M. S. Allman}
\author{R. Ancona}
\author{L. Ascarrunz}
\author{C. Barnes}
\author{J. Bartolotta}
\author{B. Bjork}
\author{P. Blanchard}
\author{M. Bohn}
\author{J. G. Bohnet}
\author{N. C. Brown}
\affiliation{Quantinuum, 303 S. Technology Ct., Broomfield, CO 80021, USA}
\author{N. Q. Burdick}
\affiliation{Quantinuum, 1985 Douglas Dr.~N., Golden Valley, MN 55422, USA}
\author{W. C. Burton}
\author{S. L. Campbell}
\author{J.~P.~Campora III}
\affiliation{Quantinuum, 303 S. Technology Ct., Broomfield, CO 80021, USA}
\author{C. Carron}
\affiliation{Quantinuum, 12001 State Hwy 55, Plymouth, MN 55441, USA}
\author{J. Chambers}
\author{J. W. Chan}
\author{Y. H. Chen}
\author{A. Chernoguzov}
\author{E. Chertkov}
\author{J.~Colina}
\author{J. P. Curtis}
\author{R. Daniel}
\author{M. DeCross}
\affiliation{Quantinuum, 303 S. Technology Ct., Broomfield, CO 80021, USA}
\author{D. Deen}
\affiliation{Quantinuum, 12001 State Hwy 55, Plymouth, MN 55441, USA}
\author{C. Delaney}
\author{J. M. Dreiling}
\affiliation{Quantinuum, 303 S. Technology Ct., Broomfield, CO 80021, USA}
\author{C. T. Ertsgaard}
\affiliation{Quantinuum, 12001 State Hwy 55, Plymouth, MN 55441, USA}
\author{J. Esposito}
\author{B. Estey}
\author{M. Fabrikant}
\author{C. Figgatt}
\author{C.~Foltz}
\author{M. Foss-Feig}
\author{D. Francois}
\author{J. P. Gaebler}
\author{T. M. Gatterman}
\author{C. N. Gilbreth}
\author{J. Giles}
\author{E. Glynn}
\author{A.~Hall}
\author{A. M. Hankin}
\author{A. Hansen}
\author{D. Hayes}
\affiliation{Quantinuum, 303 S. Technology Ct., Broomfield, CO 80021, USA}
\author{B. Higashi}
\affiliation{Quantinuum, 12001 State Hwy 55, Plymouth, MN 55441, USA}
\author{I. M. Hoffman}
\affiliation{Quantinuum, 303 S. Technology Ct., Broomfield, CO 80021, USA}
\author{B. Horning}
\affiliation{Quantinuum, 12001 State Hwy 55, Plymouth, MN 55441, USA}
\author{J. J. Hout}
\author{R.~Jacobs}
\author{J. Johansen}
\author{L. Jones}
\affiliation{Quantinuum, 303 S. Technology Ct., Broomfield, CO 80021, USA}
\author{J. Karcz}
\affiliation{Honeywell Aerospace, 12001 State Hwy 55, Plymouth, MN 55441, USA}
\author{T. Klein}
\affiliation{Quantinuum, 12001 State Hwy 55, Plymouth, MN 55441, USA}
\author{P. Lauria}
\author{P. Lee}
\author{D. Liefer}
\author{C. Lytle}
\affiliation{Quantinuum, 303 S. Technology Ct., Broomfield, CO 80021, USA}
\author{S. T. Lu}
\affiliation{Honeywell Aerospace, 12001 State Hwy 55, Plymouth, MN 55441, USA}
\author{D. Lucchetti}
\author{A. Malm}
\author{M.~Matheny}
\author{B. Mathewson}
\author{K. Mayer}
\author{D. B. Miller}
\author{M. Mills}
\author{B. Neyenhuis}
\author{L. Nugent}
\affiliation{Quantinuum, 303 S. Technology Ct., Broomfield, CO 80021, USA}
\author{S. Olson}
\affiliation{Quantinuum, 12001 State Hwy 55, Plymouth, MN 55441, USA}
\author{J.~Parks}
\author{G. N. Price}
\author{Z. Price}
\author{M. Pugh}
\author{A. Ransford}
\author{A. P. Reed}
\author{C. Roman}
\author{M. Rowe}
\author{C.~Ryan-Anderson}
\author{S. Sanders}
\affiliation{Quantinuum, 303 S. Technology Ct., Broomfield, CO 80021, USA}
\author{J. Sedlacek}
\affiliation{Quantinuum, 1985 Douglas Dr.~N., Golden Valley, MN 55422, USA}
\author{P. Shevchuk}
\author{P. Siegfried}
\author{T. Skripka}
\author{B. Spaun}
\author{R. T. Sprenkle}
\author{R.~P.~Stutz}
\author{M. Swallows}
\author{R. I. Tobey}
\author{A. Tran}
\author{T. Tran}
\affiliation{Quantinuum, 303 S. Technology Ct., Broomfield, CO 80021, USA}
\author{E. Vogt}
\affiliation{Honeywell Aerospace, 12001 State Hwy 55, Plymouth, MN 55441, USA}
\author{C. Volin}
\author{J. Walker}
\author{A. M. Zolot}
\author{J. M. Pino}
\affiliation{Quantinuum, 303 S. Technology Ct., Broomfield, CO 80021, USA}

\begin{abstract}
We describe and benchmark a new quantum charge-coupled device (QCCD) trapped-ion quantum computer based on a linear trap with periodic boundary conditions, which resembles a race track.
The new system successfully incorporates several technologies crucial to future scalability, including electrode broadcasting, multi-layer RF routing, and magneto-optical trap (MOT) loading, while maintaining, and in some cases exceeding, the gate fidelities of previous QCCD systems. The system is initially operated with 32 qubits, but future upgrades will allow for more.  We benchmark the performance of primitive operations, including an average state preparation and measurement error of 1.6(1)$\times 10^{-3}$, an average single-qubit gate infidelity of $2.5(3)\times 10^{-5}$, and an average two-qubit gate infidelity of $1.84(5)\times 10^{-3}$. The system-level performance of the quantum processor is assessed with mirror benchmarking, linear cross-entropy benchmarking, a quantum volume measurement of $\mathrm{QV}=2^{16}$, and the creation of 32-qubit entanglement in a GHZ state. We also tested application benchmarks including Hamiltonian simulation, QAOA, error correction on a repetition code, and dynamics simulations using qubit reuse. We also discuss future upgrades to the new system aimed at adding more qubits and capabilities.
\end{abstract}

\maketitle

\section{Introduction}
Several technology platforms are viable candidates for large-scale quantum computation, including trapped ions~\cite{Bruzewicz19}, neutral atoms~\cite{Henriet20}, and superconducting circuits~\cite{Kjaergaard20}. However, existing demonstrations face scaling challenges to achieve the qubit numbers and fidelities necessary for fault-tolerant quantum computing. In addition, all platforms need refinement in reliability, power consumption, form factor, and cost. This concept, known as Rent’s rule, has been discussed rigorously in terms of classical computing technologies and recently generalized to include quantum processors~\cite{FRANKE2019}. 

In this work, we characterize a trapped-ion quantum computer with a new trap design based on the QCCD architecture. The new machine, Quantinuum System Model H2, significantly increases the qubit number and decreases the physical resources per qubit, all while matching---and in some instances surpassing---the high circuit fidelity of our previous generation system~\cite{Pino2020}.

\begin{figure}[h!]
    \centering
    \includegraphics[width=0.9\columnwidth]
    {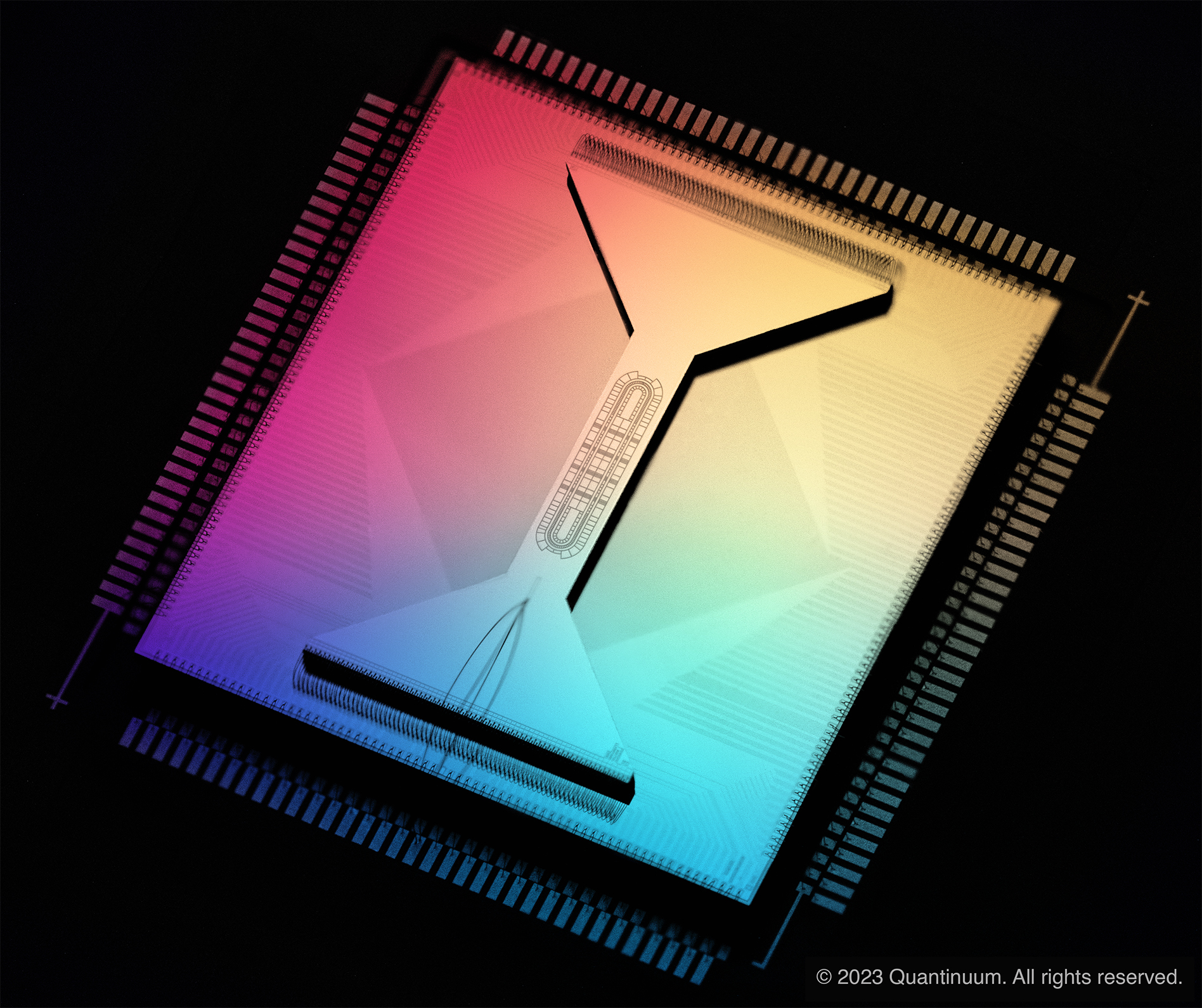}
    \caption{A picture of the H2 surface ion trap microchip. The image has been modified to enhance visibility of the trap features.  The trap sits in the isthmus in the center of the trap die.  The long axis of the trap is 6.58 mm (from the edge of the DC electrodes on either side) and the isthmus width is 2.02 mm.}
    \label{fig:prettytrappic}
\end{figure}

The QCCD architecture was proposed as a scalable method for trapped-ion quantum computation~\cite{Wineland98, Monroe2013}. Trapped-ion systems with a single trapping zone are limited in qubit number due to challenges in individually addressing single qubits within a large ion crystal, as well as motional mode crowding, which complicates achieving  high-fidelity operations in a large crystal ~\cite{CiracZollerGate,Landsman2019}.  QCCD trapped-ion systems instead have multiple trapping zones allowing operations to always be done with a small number of ions, thereby facilitating low-crosstalk addressing and maintaining high fidelity~\cite{Home2009}.  Two-qubit gates between arbitrary pairs of qubits are enabled by ion transport during a quantum circuit, which brings pairs to be gated into the same trapping zone.  Such dynamic rearrangement enables the execution of circuits with arbitrary connectivity without the overhead of logical SWAP gates typically incurred for platforms with fixed and limited connectivity~\cite{Kaushal2020}. This transport requires traps with a large number of programmable electrodes, which can be achieved using microfabricated surface traps~\cite{Seidelin2006,Maunz16, Clark2021}. 

\begin{figure*}[ht!]
\includegraphics[width=\textwidth]{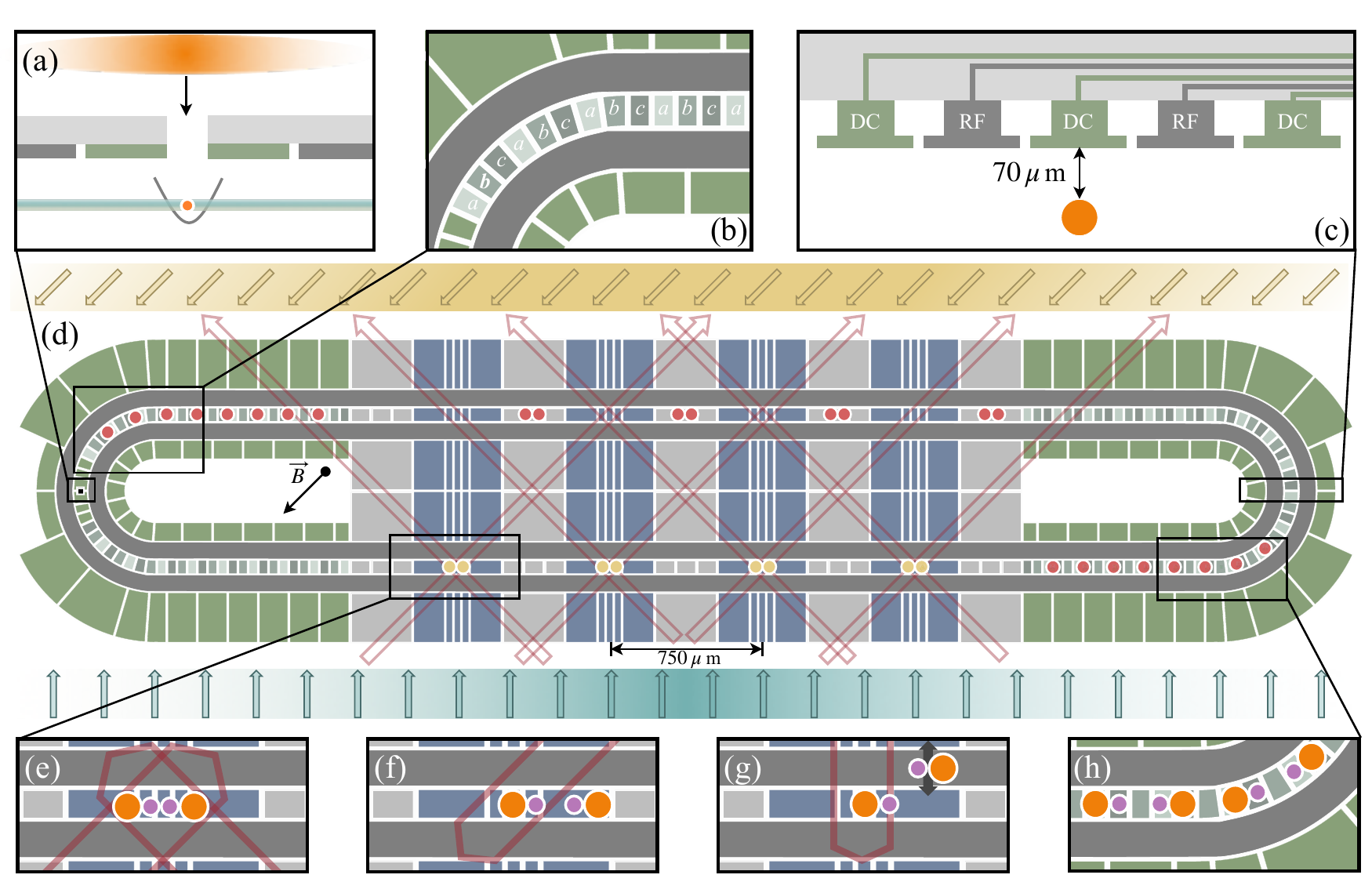}
\caption{Overview of the H2 trap including upgrades in trap design and gating operations. (a) A 2D MOT produces a collimated beam of atoms, allowing for higher neutral atom density and faster loading than an effusive oven. (b) \textit{abc} tiling of electrodes for conveyor belt transport. (c) RF tunnels to implement inner and outer RF electrodes. Ions are trapped 70 $\mu$m below the trap surface. (d) Colored top metal layer of H2 trap.  Green curved zones are conveyor belt regions for ion storage. Bottom blue zones are DG01-DG04 (from left to right), used for quantum operations. Top blue zones are UG01-UG04 gate zones (from right to left), used for sorting but not quantum operations. Darker grey loops are RF electrodes. Yellow circles represent qubits that are gated while red circles represent qubits sitting in storage during gates (note that $^{138}$Ba$^{+}$ ions are omitted for simplicity). Yellow arrows indicate the Doppler sheet beam direction while blue arrows indicate the Doppler repump sheet beam direction.  (e) Ion configuration and beam direction for 2Q gates. Large orange circles represent $^{171}$Yb$^{+}$ while smaller purple circles represent $^{138}$Ba$^{+}$. (f) Ion configuration and beam directions for 1Q  gates on left $^{171}$Yb$^{+}$. (g) Ion configuration and beam directions for state preparation and measurement (SPAM) operations on left $^{171}$Yb$^{+}$ with micromotion hiding on right $^{171}$Yb$^{+}$~\cite{Gaebler2021}. (h) Storage ion configuration in conveyor belt region.}
\label{Daytona_fig}
\end{figure*}

Our first generation hardware, H$\emptyset$ and H1 (based around the same linear trap design) \cite{Pino2020,RyanAnderson2022} demonstrated many key components of the QCCD architecture and achieved high-fidelity gates with arbitrary two-qubit (2Q) couplings.
Since the initial operation of the linear trap, qubit number $N$ was increased fivefold, from 4 in its initial mode of operation \cite{Pino2020} to 20 in its latest \cite{RyanAnderson2022}, while 2Q gate errors were decreased by roughly a factor of 5.  By increasing both the transport speeds and the number of gate zones, the average time required to execute a layer of $N$ 1Q gates and $N$/2 2Q gates on a random pairing of all $N$ qubits was kept roughly constant as $N$ increased.

This progress notwithstanding, linear geometries pose severe scaling challenges.  The time to rearrange ions for arbitrary circuit connectivity scales poorly for the linear trap design as the number of qubits increases.  The future of QCCD systems is likely in 2D traps that offer better scaling of rearrangement times and are also well suited to many error correcting codes \cite{fowler2012surface, RyanAnderson2021}. However, 2D traps present new engineering challenges that are still under development, such as junction transport~\cite{Hensinger2006,burton2022} and signal routing under the trap top metal layer. Many other aspects of QCCD scaling still in development include coupling multiple surface trap die~\cite{Brown2016,Akhtar2022}, control of a sufficient number of electrodes, laser light generation and delivery~\cite{Niffenegger2020,Menssen2022}, and detection~\cite{Todaro2021}. Not all of these challenges will be met simultaneously, but rather advances will be inserted as they are available.

This report marks the first major trap design advancement in the H-series QCCD quantum computers. Specifically, the new trap (shown in Fig.~\ref{fig:prettytrappic}) introduces: (1) RF tunnels so that RF voltage electrodes do not need to be connected on the top surface that defines the trapping potential (Sec.~\ref{sec:trap_design}), (2) voltage broadcasting to multiple control electrodes, thereby reducing the number of independent voltage sources needed to control the device (Sec.~\ref{sec:trap_design}),
and (3) MOT loading of the trap to increase the ion loading rate, decreasing the initialization time~\cite{Bruzewicz2016} (Sec.~\ref{sec:loading_state_prep}). In addition to the upgraded system design, we also report on upgraded operations, including higher performance and more efficient gating primitives. We present detailed benchmarking of the system performance with component benchmarking in Sec.~\ref{sec:compbench}, system-level benchmarking in Sec.~\ref{sec:system-level}, and algorithmic benchmarking in Sec.~\ref{sec:algobench}. Similar to the H1 series, the configuration described in this report is only the first of the H2 series, and we expect to make significant qubit count and gate zone operation upgrades in the near future.

\section{Overview of the hardware}

\subsection{Trap design} \label{sec:trap_design}
As shown in Fig.~\ref{Daytona_fig}, H2 has a race track geometry similar to traps fabricated by other groups~\cite{Amini2010, Tabakov2015, Li2017}. Two concentric RF electrodes circumscribe the center region and are driven at $\sim 200$ V and 42 MHz, creating an RF-null $70~\mu$m from the surface where ions are trapped. RF tunnels are required for the concentric RF electrodes and allow for DC electrodes to tile the full trap perimeter shown in Fig.~\ref{Daytona_fig}c. The trap has two rows of gate zones colored in blue in Fig.~\ref{Daytona_fig}, four on the top (UG01-UG04) and four on the bottom (DG01-DG04). In this work we use both rows for ion rearrangement (physical swaps), but only the DG zones are used for quantum operations (gating, state preparation, and measurement). We plan to extend quantum operations to both rows in future work.

The ``conveyor belt'' region of the trap is colored green in Fig.~\ref{Daytona_fig}d. In this region, voltage ``broadcasting'' is used to minimize DC control signals by tying multiple DC electrodes within the trap die to the same external signal. As shown in Fig.~\ref{Daytona_fig}b, each conveyor belt region contains equally spaced and sized electrodes tied together in a repeating fashion ($\{a,b,c,a,b,c,...\}$).  This requires only three total voltage signals and can support 20 wells on each side (one for every three electrodes). Additional electrodes, called shim electrodes, are located outside of the RF electrodes and used to compensate micromotion and rotate the trap principal axes. The load hole, visible in the middle of the left-side conveyor belt region in Fig.~\ref{Daytona_fig}d, is surrounded by electrodes with independent signals.

The gate zone electrode configuration is similar to that in Ref.~\cite{Pino2020} and the spacing between gate zones remains the same (750 $\mu$m).  An additional improvement to the signal count was realized by reducing the number of electrodes in the auxiliary regions around the gating zones (light grey in Fig.~\ref{Daytona_fig}d). As expected, the linear transport through the auxiliary zones is not degraded compared with H1.

In total, the trap has 376 electrodes connected to 268 independent voltage sources and 1 RF drive. Similar to H1, H2 uses a 280 pin ceramic pin grid array to connect the trap electrodes to the DC control signals. This is a reduction in the number of electrical feedthroughs per qubit in the system, which is an important metric as the number of qubits grows.

\subsection{Ion loading and state preparation} \label{sec:loading_state_prep}
H2 uses a 2D MOT as a source for neutral atoms instead of an effusive atomic oven~\cite{Johansen2022}. The MOT is connected to the main vacuum chamber via a differential pumping tube. The MOT cools both $^{171}$Yb and $^{138}$Ba neutral atoms, which are directed toward the backside load hole in the main vacuum chamber. A fraction of the neutral atoms that pass through the load hole are ionized by the photo-ionization beams on the front side, and subsequently cooled after loading into the trap (see Fig.~\ref{Daytona_fig}a). In the best case, we load one $^{171}$Yb$^{+}$ in $\sim 1.2$ ms and one $^{138}$Ba$^{+}$ in $\sim 40$ ms; however, algorithmic latency and validation procedures limit the time to load the full trap (32 $^{171}$Yb$^{+}$-$^{138}$Ba$^{+}$ [YB] pairs in a deterministic orientation) to about 3-4 minutes.
Under normal operating conditions, we observe no impact of the behavior of the quantum processor with the MOT beam on.  Once the trap is fully loaded, we detect individual loss events and replace the affected ion pairs, requiring only 10-15 seconds per lost pair. 

The qubit subspace occupies the hyperfine approximate clock states of $^{171}$Yb$^{+}$ in the $^2S_{1/2}$ state, $\ket{0} \equiv \ket{F=0, m_f =0}$ and $\ket{1} \equiv \ket{F=1, m_F=0}$.  The quantization axis is set by an externally applied magnetic field 
in the plane of the trap at 45$^{\circ}$ with respect to the long axis.  After loading, qubits are prepared in the $|0 \rangle$ state via optical pumping, similar to previous work~\cite{Olmschenk2007,Pino2020}. State preparation is currently only possible in the DG zones, so we prepare 8 qubits at a time and prepare all 32 qubits in four rounds.

\subsection{Quantum gates}
Quantum gates are implemented in the DG gate zones by stimulated Raman processes described in Ref.~\cite{Pino2020} with a laser geometry shown in Fig.~\ref{Daytona_fig}d-f. Single-qubit (1Q) gates use co-propagating beams (Fig.~\ref{Daytona_fig}f) and 2Q gates use pairs of beams with $\Delta \overrightarrow{k}$ coupling to the axial mode of motion (Fig.~\ref{Daytona_fig}e). 2Q gates are implemented with a phase-sensitive M\o lmer-S\o rensen (MS) gate~\cite{Sorensen00,Lee_2005} sandwiched between 1Q wrapper pulses (using the same laser beams as the MS interaction) to generate the parameterized gate $U_{ZZ}(\theta)=\exp{(-i\theta ZZ/2)}$~\cite{Lee_2005,Self2022,Chertkov2022, DeCross2022}.  
The value of $\theta$ is controlled by varying the detuning, duration, and Rabi rate of the MS interaction. Modeling
supports an average gate infidelity that decreases roughly linearly with $\theta$ down to a finite offset of $\approx 5\times 10^{-4}$ as $\theta\rightarrow 0$ as shown in Fig.~\ref{fig: fid versus angle}. The finite offset at zero angle exists because the wrapper pulses still occur with a delay between them, and some fraction of the 2Q laser light remains on at zero angle, leading to residual errors predominantly from laser phase noise and spontaneous emission.

The 2Q beams have the strictest requirements and consume a large portion of the total laser power budget, with the current configuration using four 2Q laser beam pairs to operate four gate zones.  We note that adding one more pair of beams would enable the operation of four more gate zones on the other side of the trap, an upgrade we plan to explore in future work.

\subsection{Measurement}
Measurement operations are performed in the DG zones with resonant beams traveling perpendicular to the long axis of the trap using state-dependent resonance fluorescence shown in Fig.~\ref{Daytona_fig}g. A photomultiplier tube array allows independent detection in all eight gate zones simultaneously, though
we only implement measurement operations in the DG zones.

Similar to previous work~\cite{Pino2020,RyanAnderson2022}, qubit measurement and reset may be performed in the middle of a quantum circuit while quantum information is preserved on other qubits. Mid-circuit measurement and reset (MCMR) causes a small crosstalk error that acts on neighboring qubits due to stray light from the measurement and reset beams (see Sec.~\ref{sec:compbench} and Table~\ref{tab:RB_avgs}).  For unmeasured ions in the gate zones, this error is mitigated by the micromotion hiding technique described in Ref.~\cite{Gaebler2021} and depicted in Fig.~\ref{Daytona_fig}g. Ions in the conveyor belt regions suffer from a similar level of crosstalk errors as ions in the gate zones, although we do not attempt to apply the micromotion hiding technique to them.

\subsection{Ion transport} \label{sec:ion transport}
Arbitrary qubit connectivity is achieved via physical ion transport. During 32-qubit operation, the ions can be grouped into four ``batches" of 8, with the four batches occupying the DG zones, UG zones, and each of the two storage regions as shown in Fig~\ref{Daytona_fig}d. The fundamental transport operations are similar to those in Ref.~\cite{Pino2020} and include split/combine, linear shifts, and physical swaps. A special type of linear shift for H2 is the batch shift, which shuttles batches of ions collectively to different regions of the trap. This operation is comparatively slow and dominates the circuit time.  The fraction of total circuit time taken up by transport varies from circuit to circuit but is 60\% on average (see Table~\ref{tab:circuit_times}).

During ion transport, we cool all $^{138}$Ba$^{+}$ ions with Doppler cooling ``sheet beams", illustrated on top and bottom of Fig.~\ref{Daytona_fig}d, which resemble sheets of laser light that cover the entire trap. These sheet beams have about a 25\% variation in intensity between the center of the trap and the edges, which does not present any performance limitations.

A compiler generates a schedule of quantum gates and transport operations with the goal of minimizing the total transport time required to execute the circuit. The circuit is first decomposed into layers, which are built iteratively by looking ahead through the circuit and grouping together (into one layer) the largest possible set of 2Q gates subject to two constraints: (1) no ions participate in more than one gate in each layer, and (2) the time ordering of 2Q gates that share one or more qubit (or any time-ordering enforced by an explicitly requested barrier) is respected.  The circuit is then converted into a layered directed acyclic graph, and a modified Sugiyama algorithm~\cite{Sugiyama81} is applied to iteratively sort the qubits in each layer of the graph in an effort to minimize the overall transport time required to execute all layers. The periodic boundary conditions of the device are explicitly taken into account and the resulting transport operations are computed using a parallel bubble sort routine that allows qubits to move in both directions around the device.

\begin{table*}[]
\begin{ruledtabular}
\begin{tabular}{lcccccc}
Circuit name                & \begin{tabular}[c]{@{}c@{}}Num. \\ ~qubits~\end{tabular} & \begin{tabular}[c]{@{}c@{}}Circuits budget (\%)\\ ~(quantum ops./transport/cooling)~\end{tabular} & \begin{tabular}[c]{@{}c@{}}Shot time \\ (s)\end{tabular} & \begin{tabular}[c]{@{}c@{}}~Num. 2Q~ \\ gates\end{tabular} & \begin{tabular}[c]{@{}c@{}}Num. 2Q \\ ~gate rounds~\end{tabular} & \begin{tabular}[c]{@{}c@{}}Num. \\ ~measurements~\end{tabular} \\ \hline
2Q RB, $\ell$ = 128$^{*}$         & 16                                                     & 2/30/68                                                                            & 1.74          & 813                                                      & 287                                                            & 16                                                            \\
Transport 1Q RB, $\ell$ = 64 & 32                                                     & 1/73/26                                                                            & 4.01          & 0                                                        & 0                                                              & 32                                                           \\
MB, $\ell$ = 10              & 32                                                     & 1/67/32                                                                            & 1.05          & 320                                                      & 80                                                             & 32                                                           \\
QV                          & 16                                                     &  1/52/47                                                                                   & 1.23          &  310                                                     & 128                                                             & 16                                                           \\
RCS                         & 32                                                     & 1/67/32                                                                           & 0.72          & 172                                                      & 56                                                             & 32                                                           \\
GHZ                         & 32                                                     & 2/64/34                                                                             & 0.18          & 31                                                       & 14                                                             & 32                                                           \\
TFIM, $Jt=7$          & 32                                             &     1/48/51                                                                                & 0.59          & 288                                                      & 72                                                             & 32                                                           \\
QAOA, $p$ = 2                  & 32                                                     & 1/69/30                                                                             & 0.97          & 96                                                       & 66                                                             & 32                                                           \\
QAOA w/qubit reuse, $p$ = 1           & 32                                                     & 1/74/25                                                                            & 2.83          & 195                                                      & 166                                                            & 130                                                          \\
Phase flip rep. code, $SE=9$            & 32                                                     & 2/57/41                                                                             & 5.4         & 540                                                     & 540                                                           & 301      \\
HoloQUADS $t$ = 24            & 32                                                     & 1/67/32                                                                             & 20.6         & 2130                                                     & 1629                                                           & 129                                                          \\ 
\end{tabular}
\end{ruledtabular}
\caption{Example circuit resource estimates. Circuit time budgets are estimated from compiler information and broken into quantum ops.~(e.g., 1Q and 2Q gates and SPAM), transport (e.g., physical swaps and shifts), and cooling (sideband cooling before 2Q gates). Shot time does not include overheads such as postchecks of ion crystal configurations, but does include the overhead for state preparation and initial cooling of about 17 ms/shot for a 32-qubit circuit.  Num.~2Q gate rounds is the number of parallel 2Q gate operations with up to four 2Q gates per round.  $^{*}$The 2Q RB is performed on 8 qubits, and an additional 8 qubits are required for the leakage gadget.}
\label{tab:circuit_times}
\end{table*}

After compilation, the layers of the circuit are then executed sequentially, with transport primitives used to arrange the ions so that qubits scheduled to be gated in a given layer are positioned next to each other.  Once arranged, we perform gates on each batch of ions, starting with the qubits already in the DG zones. Ions are transported to the center of the gate zones for quantum operations. 1Q gates are performed after moving a single YB pair into the center of the gate zone with shift operations (Fig.~\ref{Daytona_fig}f), while 2Q gates are performed with two YB pairs combined into a single four-ion crystal YBBY (Fig.~\ref{Daytona_fig}e). Before 2Q gating operations, we apply resolved sideband cooling to the $^{138}$Ba$^{+}$ ions in the DG zones~\cite{Pino2020,Monroe1995,Barrett2003}. Ions in the UG zones are transported to the nearby auxiliary zones so that they are not addressed by the gating laser beams (see red circles in Fig.~\ref{Daytona_fig}d). 
 After the gates are applied, we perform batch shifts to move new batches of ions into the DG zones and repeat the gating procedure until the full layer is completed.

A spatial phase tracking routine accounts for inhomogeneities in the magnetic field and spatially-dependent AC Zeeman shifts from the RF current~\cite{Harty2013}, which lead to spatially varying qubit frequencies. The routine calculates extraneous phase shifts that each qubit accumulates throughout the quantum circuit and compensates for them by adjusting the phase of 1Q operations appropriately. 
Imperfections in the spatial phase tracking---due to temporal instabilities in the magnetic field environment and imperfections in the calibration routines that set the 1Q optical phases---lead to memory errors during the transport operations and sideband cooling time. Additional sources of memory error are the finite T1 time of several minutes, transport failures, or background gas collisions leading to an unintentional qubit reorder (the last two are difficult to distinguish experimentally).

\subsection{Classical programming and CPU-QPU interactions}
 Quantum algorithm developers can write programs for H2 in different frameworks and languages so long as their programs compile to either OpenQASM 2.0 or QIR ~\cite{cross2017open, QIRSpec2021}. 
 Both representations contain real-time support for classical operations in the middle of the circuit, conditional expressions that rely on these classical calculations that are performed in real time, and elementary feed-forward operations conditioned on measurement results.
 
Many quantum computing applications call for interactions between classical and quantum processing units. Perhaps the most notable example is quantum error correction schemes in which syndrome measurement results are sent to a classical computer where a decoding algorithm is used to determine recovery operations and update quantum circuits in real-time. As discussed in our previous work~\cite{RyanAnderson2021,RyanAnderson2022}, we have demonstrated this capability using two different frameworks: (1) OpenQASM 2.0++, which allows for real-time decision making, and (2) a more capable classical compute environment, utilizing Web Assembly (Wasm)~\cite{wasm}, that can execute complex calculations. Option (2) has significantly enhanced capabilities aimed at the development of hybrid quantum/classical algorithms and is crucial for applications like quantum error correction.

\section{Component operations and benchmarks}\label{sec:compbench}

\begin{table}[]
\begin{ruledtabular}
\begin{tabular}{lc}
Test                  & Average infidelity ($\times 10^{-4}$) \\ \hline
1Q  RB                & 0.25(3)             \\
1Q  leakage           & 0.04(2)             \\
2Q RB                 & 18.3(5)             \\
2Q leakage            & 3.9(2)              \\
2Q SU(4) RB           & 41(1)             \\
2Q parameterized RB   & See Fig.~\ref{fig: fid versus angle} \\
Transport 1Q  RB      & 2.2(3)              \\
Measurement crosstalk & 0.045(6)            \\
Reset crosstalk       & 0.038(6)            \\
SPAM                  & 16(1)            \\
\end{tabular}
\end{ruledtabular}
\caption{Average component benchmarking results for the tests outlined in Sec.~\ref{sec:compbench}.  All values are in terms of average infidelity and $\times 10^{-4}$. The values reported here are averaged over all four zones along with the one-sigma uncertainty from semi-parametric bootstrap resampling~\cite{Meier06}.   Data from individual zones are detailed in Table~\ref{tab:rb_full_data}.}
\label{tab:RB_avgs}
\end{table}

As our first level of benchmarking, we measure the errors from various component operations in the system. Quantum operations (e.g. gates and SPAM) dominate the error budget but are only performed in the DG zones, and therefore we measure performance with a subset of eight qubits (two per DG zone). Other errors that occur during a circuit, such as memory errors, are measured with an interleaved randomized benchmarking (RB) experiment performed simultaneously on all 32 qubits. Details of each component benchmarking experiment are given below:
\begin{itemize}
\item \textbf{SPAM experiment}: Prepare each qubit in the DG zones in $\ket{0}$ and measure the probability of finding $\ket{1}$. Repeat for preparation in $\ket{1}$ and measure the probability of finding $\ket{0}$. The average is the SPAM error per qubit. 
This procedure cannot differentiate between state preparation and measurement errors; however, detailed modeling 
predicts that the SPAM error is dominated by measurement error for $^{171}$Yb$^{+}$~\cite{Acton2006,Noek13}.

\item \textbf{1Q gate randomized benchmarking (1Q  RB)}: We use the standard Clifford-twirl randomized benchmarking for measuring the error of 1Q  gates~\cite{Magesan11} with a random final Pauli to fix the asymptote~\cite{Harper19}. We report the average infidelity per 1Q Clifford.
\item \textbf{2Q gate randomized benchmarking (2Q RB)}: Similar to 1Q RB, we use the Clifford-twirl technique~\cite{Magesan11} for measuring the error of 2Q gates. Each 2Q Clifford is constructed with zero to three $U_{ZZ}(\pi/2)$ gates and each sequence includes a random final Pauli to fix the asymptote~\cite{Harper19}. We scale the 2Q Clifford average infidelity by the average number of $U_{ZZ}(\pi/2)$ gates per Clifford, which is 1.5, and report that as the average infidelity per 2Q gate. An example decay plot is shown in Fig.~\ref{TQ_RB_decay}a.

\item \textbf{2Q SU(4) gate randomized benchmarking (2Q SU(4) RB)}: We use the same general technique as 2Q RB, but instead of 2Q Cliffords we use unitaries randomly sampled from the Haar measure over SU(4) constructed with three parameterized $U_{ZZ}(\theta)$ gates, and for each sequence include a random final Pauli to fix the asymptote~\cite{Harper19}. We report the average infidelity per SU(4) operation. 

\item \textbf{2Q parameterized gate randomized benchmarking}:
We use a direct randomized benchmarking procedure~\cite{Proctor2019} to measure the average infidelity of the parameterized 2Q gate $U_{ZZ}(\theta)$ as a function of angle $\theta$. The details of the protocol are in App.~\ref{app:arb_rb}. A plot of the average infidelity versus angle is shown in Fig.~\ref{fig: fid versus angle}.

\begin{figure}
\includegraphics[width=0.45\textwidth]{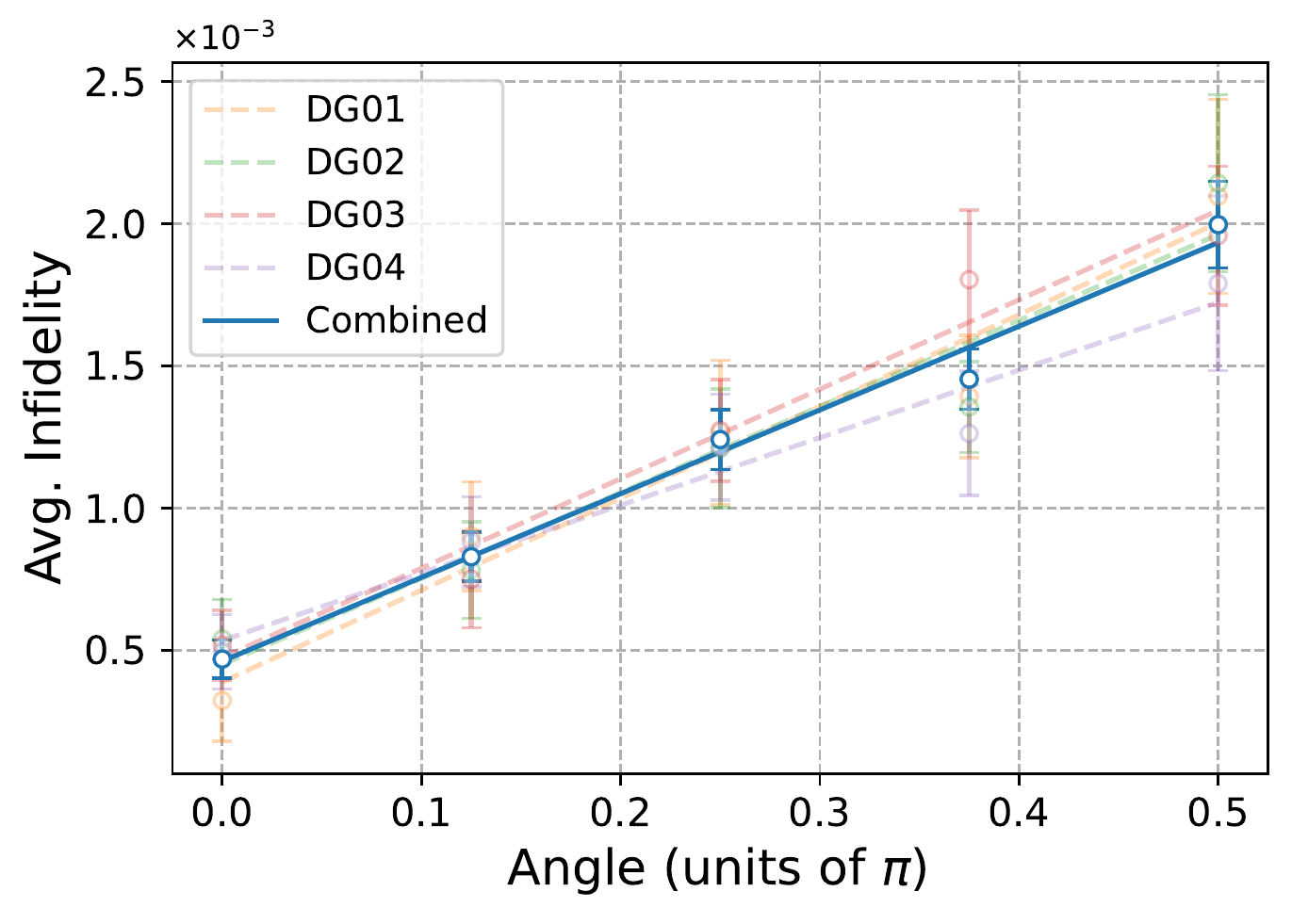}
\caption{Average infidelity as a function of angle for the parameterized 2Q gate $U_{ZZ}(\theta)$.
Each data point is obtained by fitting the decay curves shown in Fig.~\ref{fig: arbrb decay} to an exponential decay function.
The infidelity at $\theta=0$ is due to both the wrapper pulses and memory error incurred during the cooling pulses,
which are still applied in the absence of an MS gate.
The linear best-fit to the zone-averaged data is given by $\epsilon(\theta)= \big(2.9(2) \, \theta/\pi+0.46(6)$\big) $\times10^{-3}$.}
\label{fig: fid versus angle}
\end{figure}

\item \textbf{Measurement/reset crosstalk depumping}: Measurement/reset crosstalk errors are estimated with bright-state depumping experiments~\cite{Gaebler2021} where a subset of qubits are prepared in $\ket{1}$ and other qubits are measured/reset repeatedly. The qubits in $\ket{1}$ decay due to crosstalk errors from the repeated process, and the decay rate scales with the average infidelity. 

\item \textbf{Interleaved transport randomized benchmarking (Transport 1Q RB)}: During a circuit, qubits incur errors in between successive 2Q gates due to idling during transport/cooling (memory errors) and the application of 1Q gates (errors from any mid-circuit measurements or resets are considered above).  The contribution of these errors to representative circuits is measured with an interleaved 1Q RB experiment on all 32 qubits: 1Q Clifford gates are interleaved with ``dummy" 2Q gates on 16 random pairings (our choice for a representative transport sequence). The dummy 2Q gates force ion rearrangement, transport for 2Q gating, and sideband cooling but do not apply any 2Q lasers, avoiding 2Q gate errors and leaving the 32 qubit state fully separable. The resulting RB decay rate depends on both the average 1Q gate infidelity and memory errors, but we expect that memory errors are the dominant contribution.
\end{itemize}
Additional experimental details and data can be found in App.~\ref{app:RB_data}. Results from these experiments are reported in both Table~\ref{tab:RB_avgs} (averaged over zones) and Table~\ref{tab:rb_full_data}. An example breakdown of circuit timing for 2Q RB and transport 1Q RB is shown in Table~\ref{tab:circuit_times}.

\begin{figure}
\includegraphics[width=0.45\textwidth]{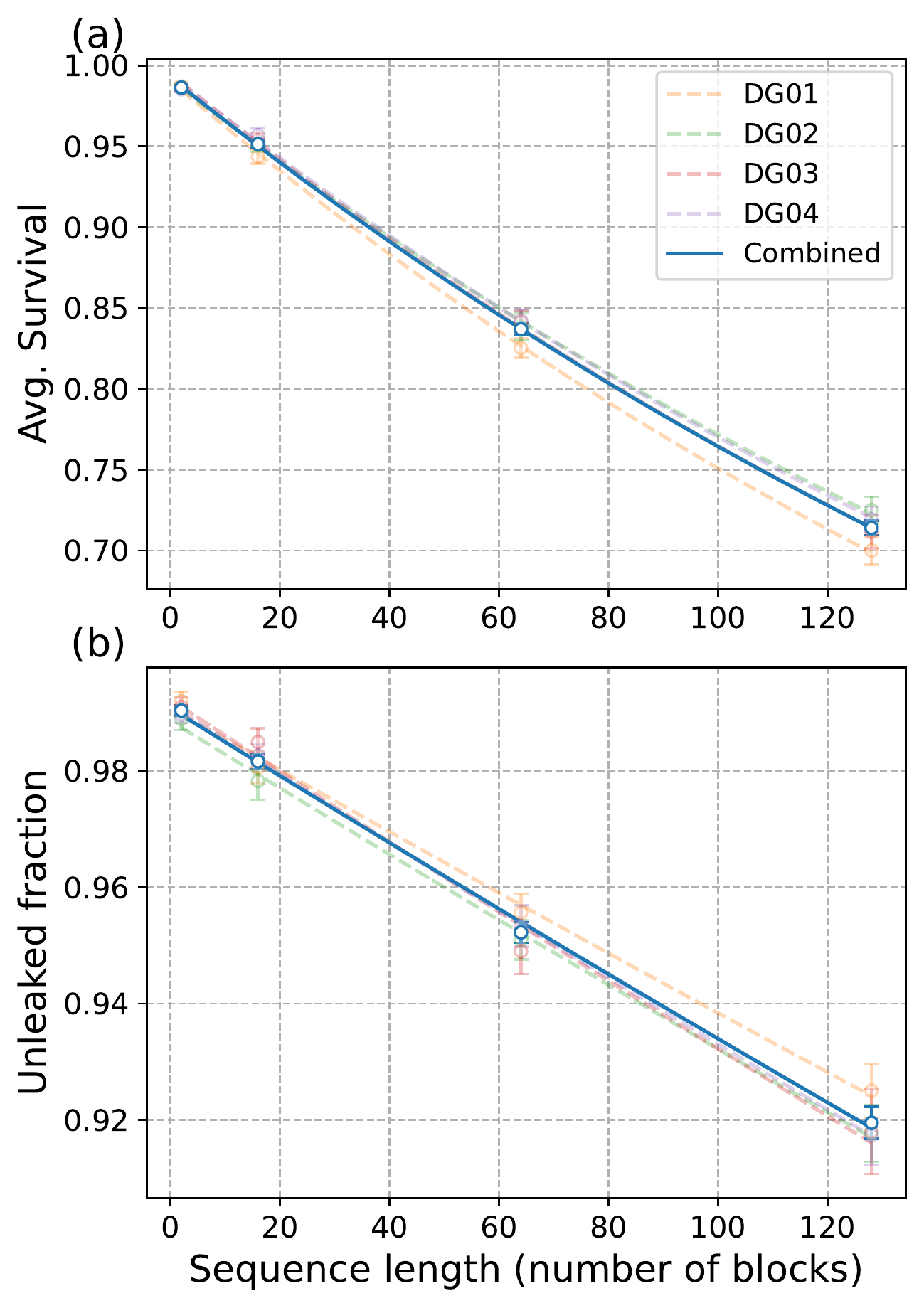}
\caption{2Q randomized benchmarking decay curves for each zone and for the combined average across all zones. (a) Standard RB decay curve.  The average infidelity per 2Q gate is 1.83(5)$\times10^{-3}$ across all four gate zones.  (b) Decay of fraction of shots without leakage on either qubit as identified by the leakage detection gadget, which gives a measured leakage rate per 2Q gate of 3.9(2)$\times10^{-4}$ across all four gate zones.
}
\label{TQ_RB_decay}
\end{figure}

\begin{figure}
\begin{center}
\includegraphics[width=0.4\textwidth]{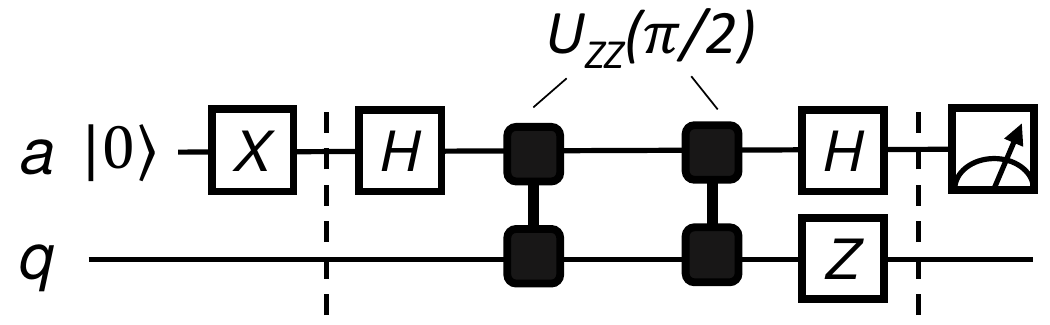}
\end{center}
\caption{Leakage detection gadget, adapted from Ref.~\cite{Stricker2020}.
The gadget uses an ancilla qubit `$a$' to detect whether qubit `$q$' has leaked.
The ancilla is initially prepared in $\ket{1}$.
If `$q$' has leaked, the 2Q gates have
no logical effect and `$a$' is measured as $\ket{1}$.
If `$q$' has not leaked, then the gadget (within the barriers in the circuit diagram)
acts as $X_aI_q$, and
`$a$' is measured as $\ket{0}$.
}
\label{leak_det_fig}
\end{figure}

For 1Q and 2Q RB, we also measured the rate of leakage errors per gate by applying a ``leakage detection gadget" at the end of each circuit, as illustrated in Fig.~\ref{leak_det_fig}. The leakage detection gadget uses an ancilla qubit to flag shots that had a leakage error, i.e., an error that moved population outside of the computational subspace. In our system, leakage is most likely due to the unavoidable spontaneous emission that occurs in gates driven by a stimulated Raman process. The leakage rate per gate $r_L$ is defined as the rate that population leaves the computational subspace (whether 1Q or 2Q). We can estimate $r_L$ by repeating the gate $\ell$ times, applying the gadget to each gated qubit, and fitting the leakage detection rate as shown in Fig.~\ref{TQ_RB_decay}b. Further details are given in App.~\ref{app:leakage}.

\section{System-level benchmarks}\label{sec:system-level}
Benchmarks of component operations are a crucial fine-grained tool for estimating the contribution of various errors to quantum circuits. However, there are many potential ways in which they can mischaracterize device performance, for example when crosstalk or non-Markovian errors are present. Therefore, it is important to also benchmark performance on a variety of more complex, multi-qubit circuits, and to assess to what extent that performance can be understood from the measured performance of the constituent operations. Here we present results from four system-level benchmarks: (A) mirror benchmarking~\cite{Proctor2022,Mayer2021}, (B) quantum volume (QV)~\cite{Cross2019,Baldwin2022}, (C) linear cross-entropy measurements for random 2D circuits \cite{google_supremacy_2019}, and (D) creation/certification of N-partite entanglement in GHZ states.

In benchmarks (A-C), the random structure of the circuits justifies simple heuristic arguments relating the overall circuit performance to the component operation fidelities.  In each case, we assume that all non-SPAM errors can be attributed to the 2Q gates themselves, and come in the form of a depolarizing channel (with uniform fidelity) attached to each 2Q gate.  This approach accumulates all errors that happen to qubits between 2Q gates, primarily due to SQ gate errors and memory errors, and lumps them in with the 2Q gate to form an effective ``per 2Q gate'' error rate, which we denote by $\epsilon^{\rm 2Q}_{\rm eff}$.  To obtain a simple but reasonable estimate for $\epsilon^{\rm 2Q}_{\rm eff}$ based on the component benchmarks, we first determine the average angle $\bar{\theta}$ of 2Q gates used in the system-level benchmark. The data and linear fit reported in Fig.\,\ref{fig: fid versus angle} 
then allows us to estimate the 2Q gate contribution. 
We then add this 2Q gate contribution together with twice the error from Transport 1Q RB, 
giving a predicted effective error
\begin{align}
\epsilon_{\rm eff}^{\rm 2Q}=10^{-3}\big(2.9(2)\bar{\theta}/\pi+0.9(1)\big).
\label{eq:2Qeff}
\end{align}
Using analyses described in the appendices, we also extract an inferred $\epsilon_{\rm eff}^{\rm 2Q}$ from the system-level benchmarking data presented below, and report the comparisons to the predicted values in Table~\ref{tab:2Qeff}.  The agreement is not perfect, nor is it expected to be, given that memory errors can be highly circuit dependent.  For example, in QV circuits, multiple 2Q gates happen with very little delay in between; whereas, the memory error per 2Q gate inferred from Transport 1Q RB assumes a single random reconfiguration of ions between every repeated gate on a given qubit, which likely contributes to the overestimate of $\epsilon_{\rm eff}^{\rm 2Q}$ reported in Table~\ref{tab:2Qeff}.  Nevertheless, the overall reasonable agreement between predicted and inferred values suggests that the results of large-scale circuits are generally well aligned with expectations based on the individual component benchmarks.
\begin{table}[]
\begin{ruledtabular}
\begin{tabular}{lccc}
Source & $\bar\theta$ & \begin{tabular}{c}$\epsilon_{\rm eff}^{\rm 2Q}$ \\ (inferred)\end{tabular} & \begin{tabular}{c}$\epsilon^{\rm 2Q}_{\rm eff}$  \\(predicted)\end{tabular} \\ \hline
Mirror benchmarking & $0.5\pi$ & 2.6(2) & 2.4(1) \\
Quantum volume & $0.35\pi$ & 1.7(1)  & 1.9(1) \\
Random circuit sampling~~~ & $0.42\pi$ & 1.9(2)   & 2.1(1)\\
\end{tabular}
\end{ruledtabular}
\caption{\label{tab:2Qeff}
Effective error per 2Q gate $\epsilon_{\rm eff}^{\rm 2Q}$ ($\times 10^{-3}$) inferred from system-level benchmarks, compared to the average per 2Q gate error estimated from combining the component benchmarks as described in the text [see Eq.\,(\ref{eq:2Qeff})].
}

\end{table}

\subsection{Mirror benchmarking}

Circuit mirroring was introduced as a scalable way
to benchmark arbitrary quantum circuits~\cite{Proctor2022, Proctor2022_2}.
We perform a randomized circuit mirroring experiment that we refer to as mirror benchmarking (MB).
As described in Ref.~\cite{Mayer2021}, MB circuits consist of layers of 1Q gates on all qubits and 2Q gates between random pairings of the qubits with full connectivity. The 1Q gates are Clifford gates sampled uniformly at random, and each 2Q gate is the native $U_{ZZ}(\pi/2)$ gate. The circuits are ``mirrored", meaning that the inverse circuit is applied in the second half. A final random $N$-qubit Pauli is applied to randomize the ideal outcome for each circuit. The circuits also employ Pauli randomization on the 2Q gates so that the error channel per layer can be treated as stochastic Pauli~\cite{Wallman2016}. The circuit-averaged probability of observing the ideal outcome as a function of the number of circuit layers will then decay exponentially.
If the 2Q gate error channel is depolarizing, then the decay parameter as a function of the 2Q gate average fidelity is given by an analytic formula (Eq.~C4 in Ref.~\cite{Mayer2021}). Fitting experimentally measured decay curves to exponentials and inverting this formula provides an effective 2Q infidelity for the system that includes 1Q  gates, 2Q gates, and the memory error for random permutations.

\begin{figure}
\includegraphics[width=0.45\textwidth]{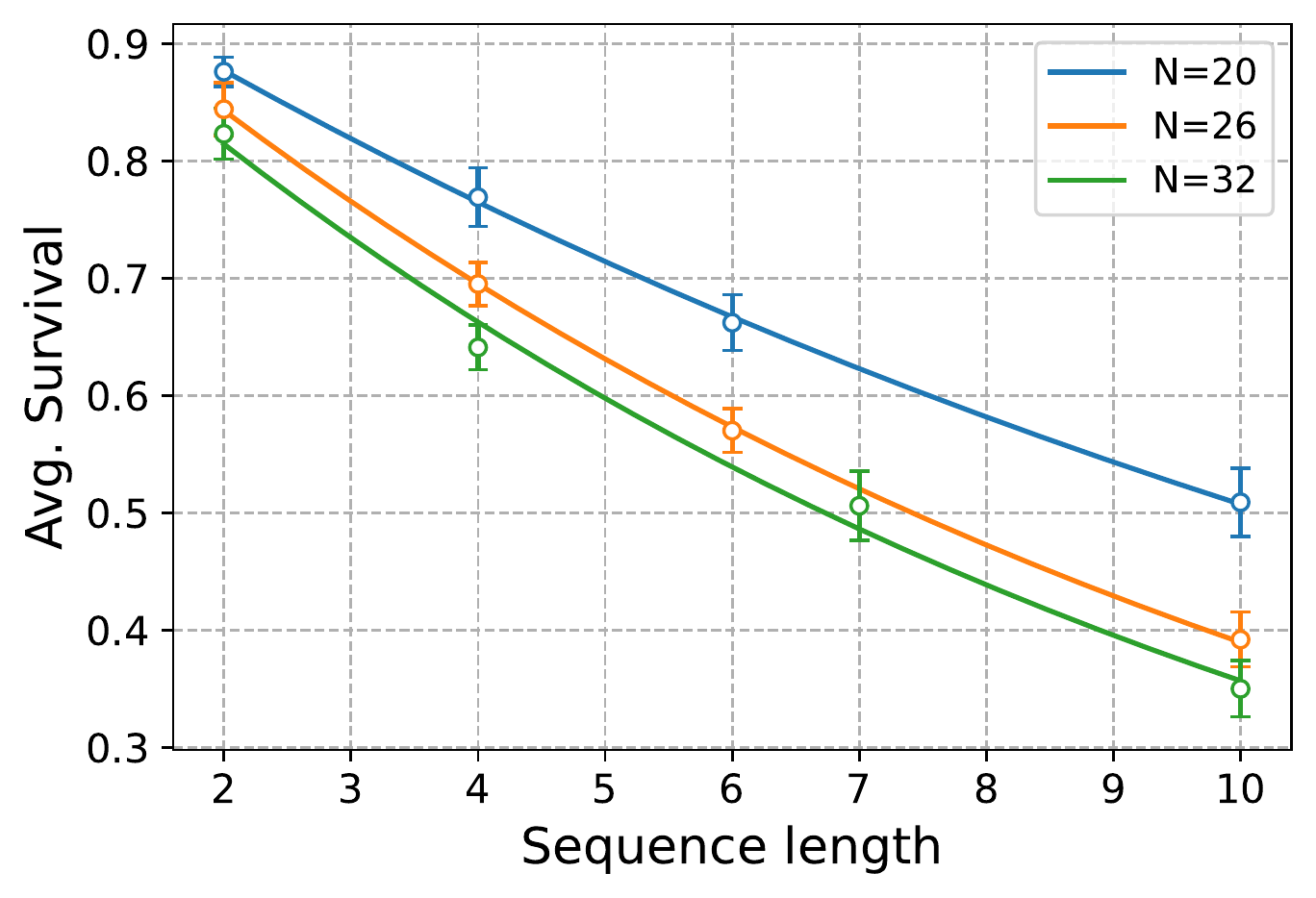}
\caption{Mirror benchmarking experiments on H2.
The sequence lengths correspond to half the circuit depths,
so circuits of length $\ell$ contain $2\ell N$ 2Q gates with full connectivity.
10 random circuits were run at each sequence length with 100 shots per circuit.
The average survival probabilities are fit to the model $p(\ell)=Au^{\ell-1}$.
The parameter $u$ is used to obtain an
effective 2Q gate average fidelity for a constant 2Q depolarizing error~\cite{Mayer2021}.}
\label{MB_fig}
\end{figure}

We performed MB experiments on H2 with $N$=20, 26, and 32 qubits.
The decay plots are shown in Fig.~\ref{MB_fig},
and the results are listed in Table.~\ref{MB_table}.
For $N=32$, we find $\epsilon_{\rm eff}^{\rm 2Q} = 2.6(2)\times10^{-3}$.
Importantly, we find that $\epsilon_{\rm eff}^{\rm 2Q}$ does not increase with qubit number.

\subsection{Quantum volume}
Quantum volume is a system-level test designed to be comparable across gate-based quantum computers. 
The QV test is run with a collection of random circuits acting on $N$ qubits. Each random circuit is generated by randomly pairing all qubits, applying random $\mathrm{SU}(4)$ unitaries
to each pair, and repeating for $N$ rounds. The performance is assessed with a heavy-output test that requires classical simulation of the quantum circuits. The test is passed when the probability of generating heavy-outputs is greater than 2/3 with two-sigma confidence, which yields a measured value of $\textrm{QV}=2^N$~\cite{Cross2019}. A totally decohered circuit  returns heavy outputs half the time, so the QV test's threshold of 2/3 requires that the errors are small enough to be strongly distinguishable from a random distribution. Therefore, a QV of $2^N$ implies high performance on many circuits with more than $N$ qubits and/or depth greater than $N$, as evidenced by several example algorithms run on all $32$ qubits in Sec.\,\ref{sec:algobench}. QV has been measured on a variety of different systems~\cite{Pelofske22} with the largest previously reported value of $\textrm{QV}=2^{15}$ from H1~\cite{qv_n=15}.

\begin{figure}
\includegraphics[width=0.45\textwidth]{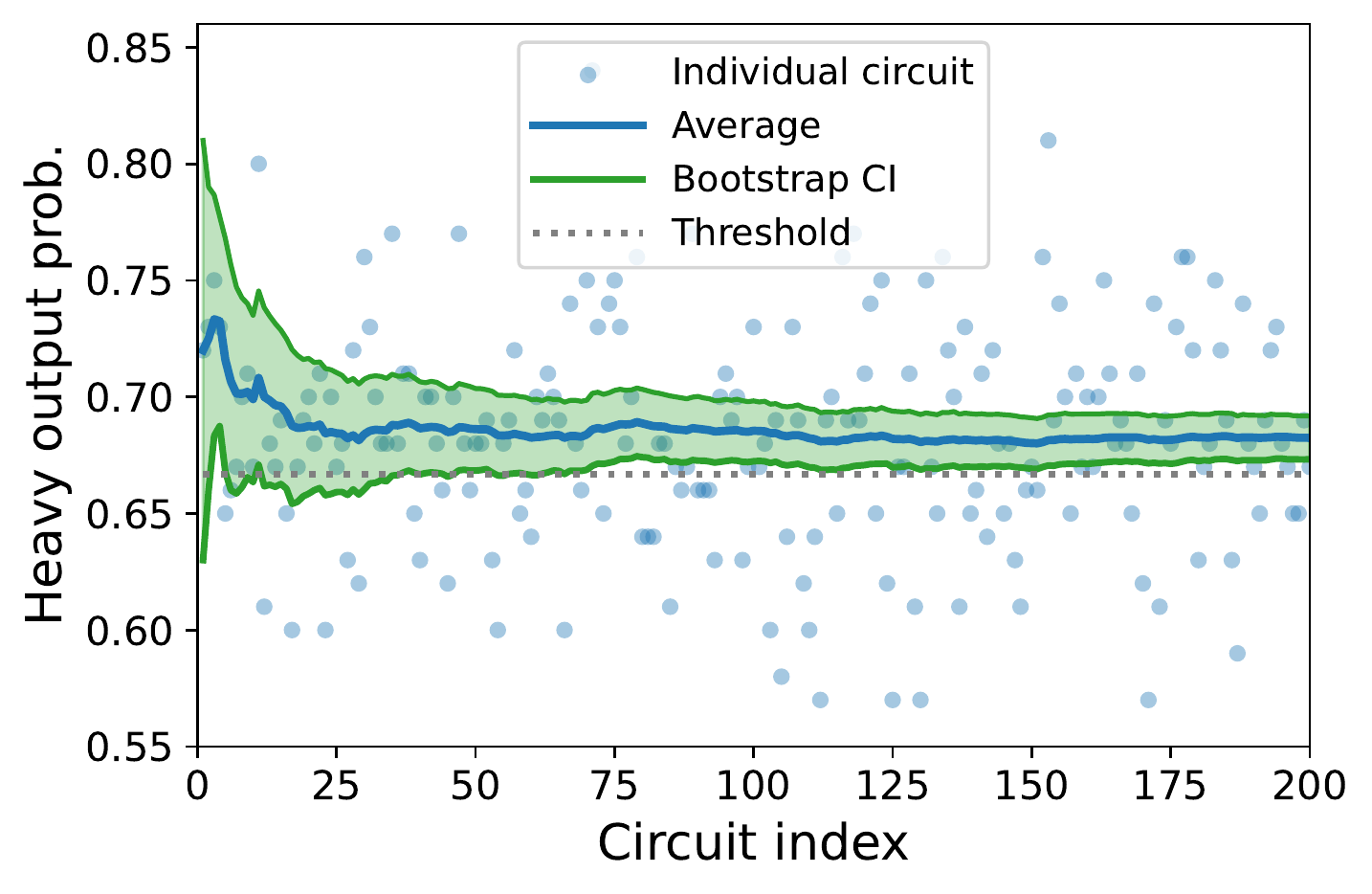}
\caption{Quantum volume $\textrm{QV}=2^{16}$ quantum volume measurement on H2. The average and two-sigma confidence interval of the heavy-output probability are plotted as a function of the circuit index. Passing occurs when the green shaded region (two-sigma confidence interval from semi-parametric bootstrap method) is above the dashed grey line at 2/3, which we satisfy in a dataset with 200 randomly generated circuits.}
\label{QV_fig}
\end{figure}

We performed several QV measurements, with the highest measured value being $\textrm{QV}=2^{16}$. The $\textrm{QV}=2^{16}$ test data is shown in Fig.~\ref{QV_fig} and used 200 randomly generated circuits each run with 100 shots and using an average of 296 parameterized 2Q gates. The measured heavy-output probability is 68.2\%, which clears the minimum threshold of 2/3 with greater than two-sigma confidence calculated by the semi-parametric bootstrap method outlined in Ref.~\cite{Baldwin2022}. 

\subsection{Random circuit sampling}
A system-level benchmark of recent interest is the computational task of sampling the output distributions of random quantum circuits (RCS). Like QV, RCS is not a scalable benchmark as it requires classical computation time exponential in $N$. It was recently proven \cite{Aharonov2022} that at fixed gate error RCS is not a scalable route to quantum supremacy at large $N$; however, it still tests the quantum computer's ability to faithfully execute circuits for which classical simulation methods are, at least in practice, extremely difficult given high enough gate fidelities. Also, it has been run on a variety of quantum computers in the context of quantum advantage demonstrations \cite{google_supremacy_2019, PhysRevLett.127.180501, Zhu21}, making it useful from the standpoint of cross-platform comparisons.

\begin{figure}[]
\includegraphics[width=0.45\textwidth]{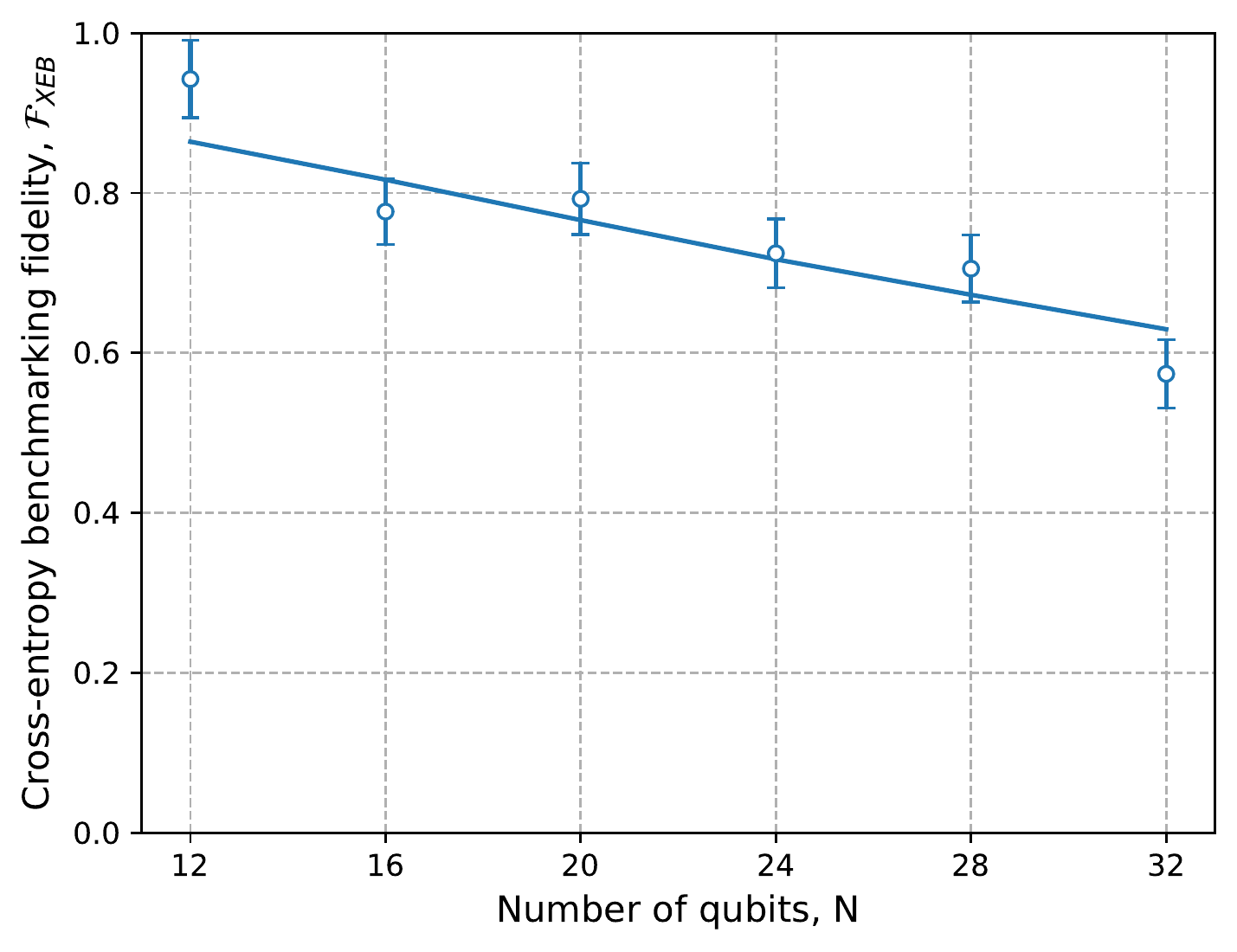}
\caption{Linear cross-entropy benchmarking fidelity as measured on H2 for classically verifiable random circuits. Each data point displays the combined results from 10 circuits each executed with 100 shots. The details of the best-fit curve are described in App.~\ref{app:rcs}.\label{fig:crossentropy}}
\end{figure}

We structure our circuits as if the qubits involved tile a two-dimensional grid with nearest-neighbor interactions~\cite{google_supremacy_2019}, although we emphasize that this constraint is only imposed for fair comparison with prior art and is not a hardware constraint of H2. Future work may study whether random circuits built from randomly gating pairs of qubits with arbitrary connectivity achieves a greater degree of classical simulation difficulty at a reduced circuit depth. At each $N$, the grid dimensions are chosen to be as close to square as possible. In each layer from the circuit, a 1Q gate chosen randomly from $\{\sqrt{X}, \sqrt{Y}, \sqrt{W}\}$, where $W = \frac{1}{\sqrt{2}} (X+Y)$, is applied to each qubit. The 1Q gate applied to a given qubit in one layer is omitted from the set of possible 1Q gates applied to the same qubit in the next layer. Subsequently, a 2Q gate is applied to pairs of qubits following a particular tiling pattern on the grid (see Ref.~\cite{google_supremacy_2019}, Fig. 3). A final round of 1Q gates is applied to all qubits before measurement. We implement the exact same 2Q gate as in Ref.~\cite{google_supremacy_2019}, 
\begin{align}
    \text{fSim} \left(\frac{\pi}{2}, \frac{\pi}{6}\right) = e^{-i\theta(X\otimes X + Y\otimes Y) / 2} e^{-i\phi(I-Z)\otimes(I-Z)/4}.
\end{align} 
Up to 1Q gates,
\begin{equation}
\begin{split}
\text{fSim} \left(\frac{\pi}{2}, \frac{\pi}{6}\right)  & \simeq \text{iSWAP} \cdot U_{ZZ}\left(\frac{\pi}{12}\right)  \\
 & \simeq  \text{SWAP} \cdot U_{ZZ}\left(-\frac{5\pi}{12}\right).
\end{split}
\end{equation}

The SWAP gate is handled in software by relabeling and transporting qubits, so the $\text{fSim} (\frac{\pi}{2}, \frac{\pi}{6})$ gate is implemented on H2 with exactly one 2Q gate. In practice, we generate the circuits using the Sycamore gate defined in the \verb|pytket| library \cite{Sivarajah_2021}; \verb|pytket|'s compilation to Quantinuum hardware automatically rebases the circuits to use a $U_{ZZ}(5\pi/12)$ gate via the above identity with the addition of 1Q gates that are absorbed into the randomly chosen 1Q gates.

Since H2 currently supports 32 qubits, well within the classically feasible regime, we focus on the ``classically verifiable" repeating EFGHEFGH gate tiling pattern from Ref.~\cite{google_supremacy_2019}. As developed in Ref.~\cite{Aaronson2021}, we use linear cross-entropy benchmarking to quantify the success of the quantum computer in sampling from the true output distribution of each random circuit. This procedure computes a quantity called the linear cross-entropy benchmarking fidelity, $\mathcal{F}_{\text{XEB}}$. To match the parameters in Ref.~\cite{google_supremacy_2019}, we explore random circuits of depth 14, and average the resulting $\mathcal{F}_{\text{XEB}}$ over 10 circuits at each fixed $N$, combining the uncertainties on each measurement of $\mathcal{F}_{\text{XEB}}$ by inverse-variance weighting. The measured results on H2 are displayed in Fig.~\ref{fig:crossentropy}. With future improvements to the number of qubits in H2, assuming comparable $\epsilon_{\rm eff}^{\rm 2Q}$, we expect the cross-entropy benchmark results will pose serious challenges to classical simulations.

\subsection{N-partite entanglement certification in GHZ states}\label{sec:GHZ}

The $N$-qubit GHZ state \cite{Greenberger1989} is defined as
\begin{equation}
    \ket{GHZ_N}=\frac{1}{\sqrt{2}}\big(\ket{0}^{\otimes N}+\ket{1}^{\otimes N}\big).
\end{equation}
Producing GHZ states is a demanding test of qubit coherence, as they are maximally sensitive probes of global dephasing. Moreover, GHZ state fidelities have been widely measured and reported across a variety of quantum hardware~\cite{Pogorelov2021, Omran2019, Song2019, Thomas2022, Mooney2021}, making this test helpful for assessing the performance of the H2 device in a broader context.

\begin{figure}
\includegraphics[width=\columnwidth]{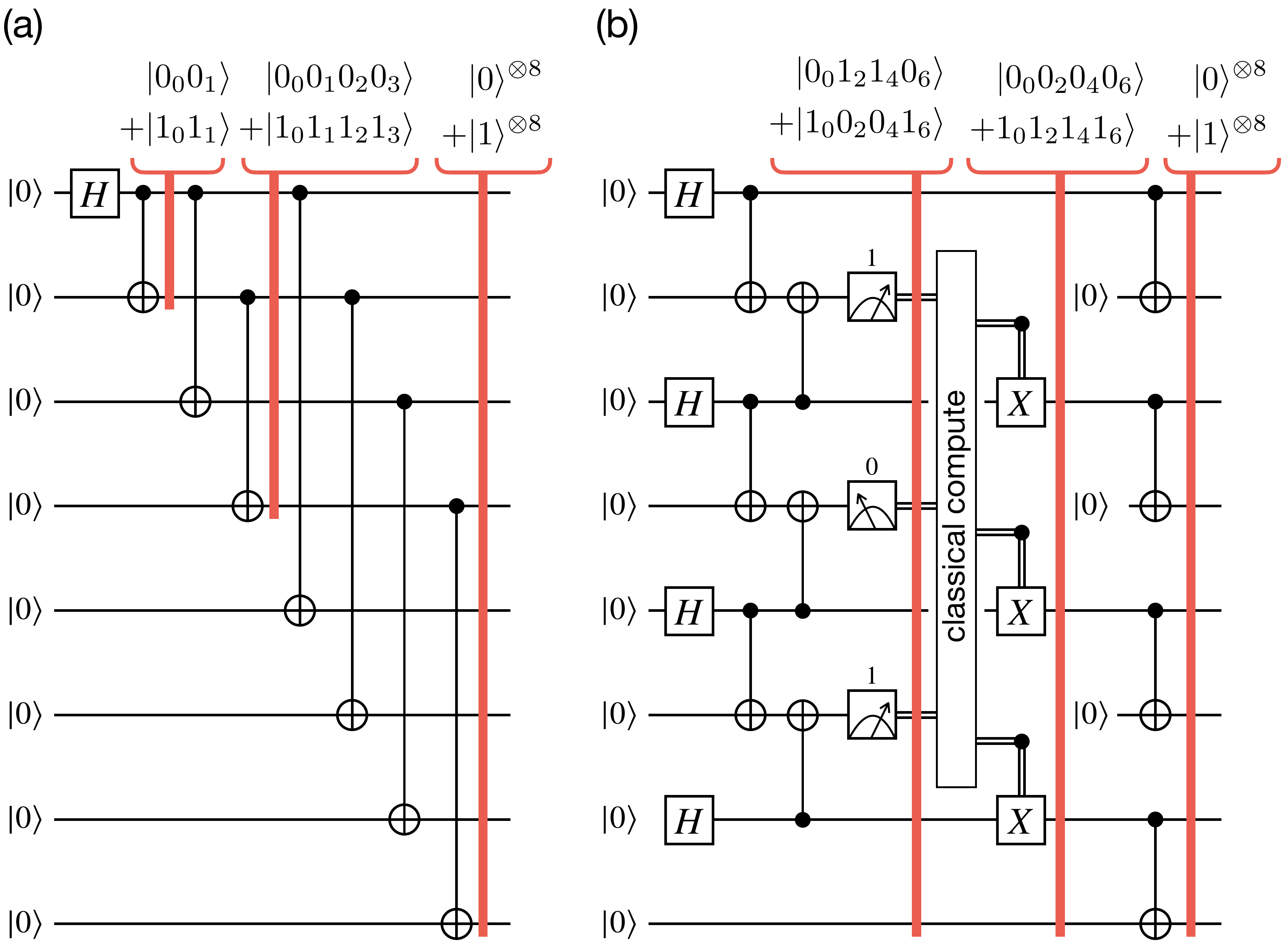}
\caption{GHZ state preparation circuits for (a) log-depth unitary and (b) constant-depth adaptive preparation, here shown for $N=8$ for simplicity.}
\label{fig:GHZ_prep}
\end{figure}

We prepare GHZ states of $N$=20, 26, and 32 using
the log-depth circuit construction given in Ref.~\cite{Cruz2019} and an $N=32$ GHZ state using a constant-depth adaptive circuit construction~\cite{Briegel2001}.
The latter was submitted via OpenQASM 2.0++ and exemplifies how mid-circuit measurement
and feed-forward can be used to create long-range-entangled states in constant depth~\cite{Lu2022, fossfeig2023,iqbal2023topological}.
Both circuit constructions are shown in Fig.~\ref{fig:GHZ_prep}.

We estimate the fidelity of the GHZ states using the
method of Ref.~\cite{Guehne2007}.
The fidelity of a density matrix $\rho$ with respect to the GHZ state is
\begin{multline}\label{eq: GHZ fidelity}
    F(\rho, \ket{GHZ_N}) =
    \frac{1}{2}\mathrm{Tr}(\rho \ket{0}\bra{0}^{\otimes N}) + \frac{1}{2}\mathrm{Tr}(\rho \ket{1}\bra{1}^{\otimes N})\\
    + \frac{1}{2}\mathrm{Tr}\big(\rho(\ket{0}\bra{1}^{\otimes N}+\ket{1}\bra{0}^{\otimes N})\big).
\end{multline}
The first two terms are the populations in the
all-zero and all-one states and are estimated by
measuring all qubits in the computational basis.
The third term is estimated using the fact that
\begin{equation}
    \ket{0}\bra{1}^{\otimes N} + \ket{1}\bra{0}^{\otimes N} = \frac{1}{N}\sum_{k=1}^N (-1)^k M_k,
\end{equation}
where the operators
\begin{equation}\label{eq: GHZ parity operator}
    M_k = \biggl(\cos(k\pi/N)X + \sin(k\pi/N)Y\biggr)^{\otimes N},
\end{equation} 
for $k\in\{1,\dots,N\}$ correspond to
the global parity of spin along
the axis $\theta_k=k\pi/N$ on the equator of the Bloch sphere,
and can be measured with only 1Q rotations.
The complete fidelity estimation protocol requires $N+1$ measurement bases.

\begin{figure*}
\includegraphics[width=0.95\columnwidth]{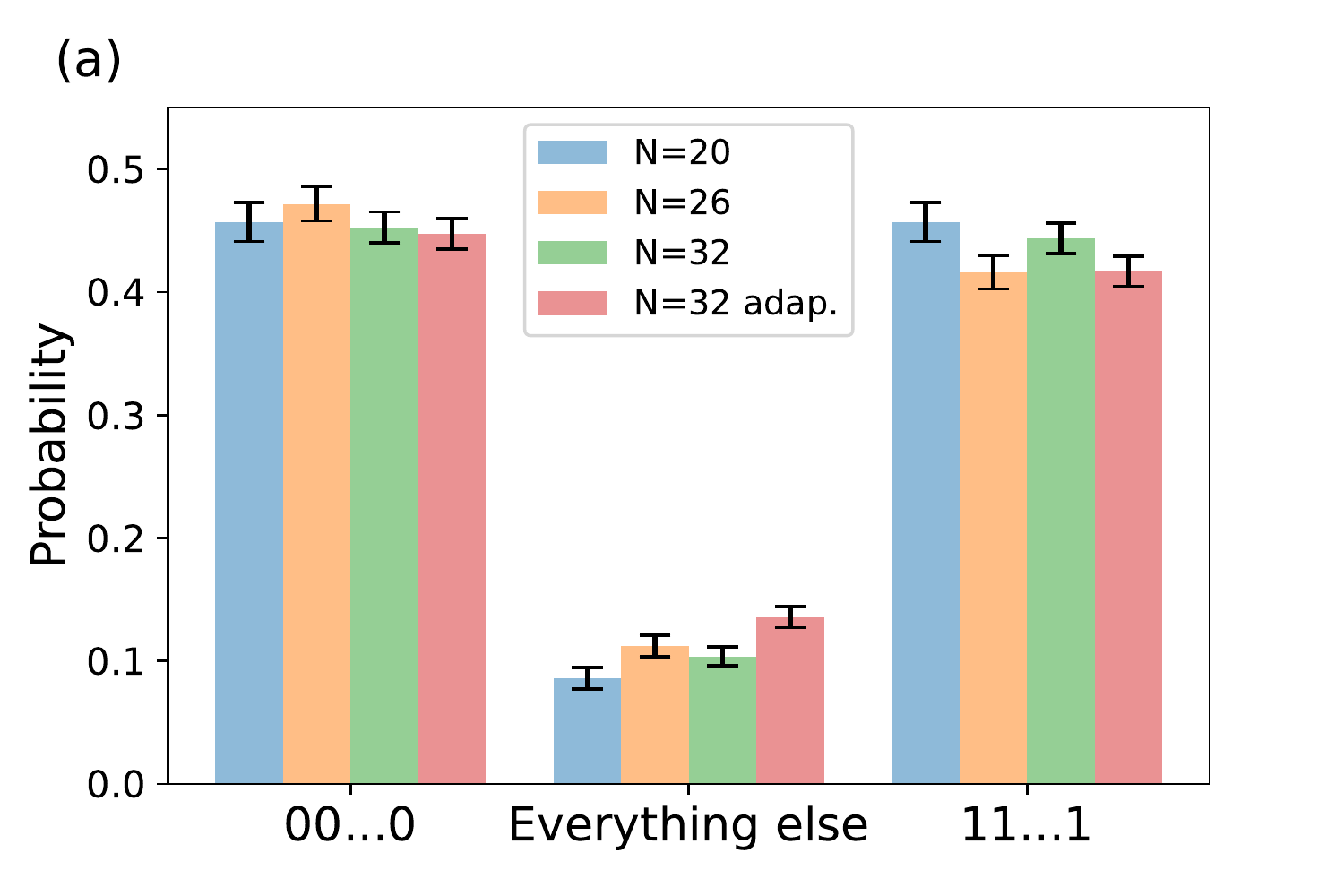}  
\includegraphics[width=0.95\columnwidth]{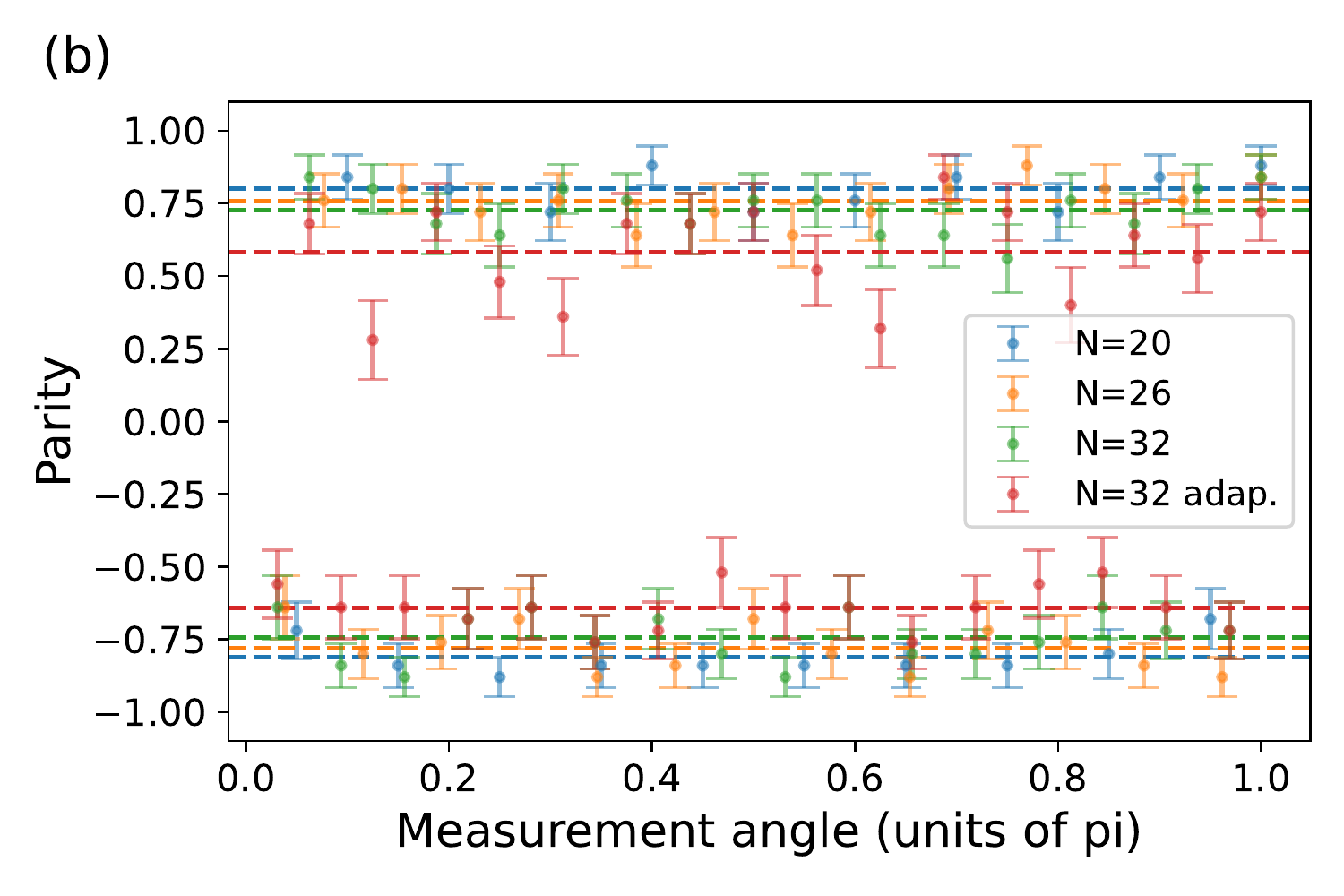}

\caption{Populations and parities of $N$=20, 26, and 32-qubit GHZ states constructed with a log-depth unitary protocol, and also of a $32$ qubit GHZ state produced with a constant-depth adaptive circuit.
(a) Populations of $\ket{0}^{\otimes N}$ and $\ket{1}^{\otimes N}$.
The ideal GHZ state has populations of $0.5$ in these two states and zero in all other states.
(b) Expectation values of the operator $M_k$ defined in Eq.~\ref{eq: GHZ parity operator}, plotted versus angle $\theta_k=k\pi/N$. The ideal GHZ state has values of $1$ and $-1$ for even and odd $k$, respectively.
The dashed lines denote the averages.}
\label{fig:GHZ_pops_parities_versus_N}
\end{figure*}

We ran one circuit with 50 shots for each of the $N$ measurements of $M_k$,
and $N$ circuits with 50 shots for the population measurements.
All the log-depth unitary preparation circuits across the 
various $N$ were run in a random order.
The results of the population and parity measurements are shown in Fig.~\ref{fig:GHZ_pops_parities_versus_N},
and the estimated state fidelities are listed in Table~\ref{GHZ_table}.
For $N=32$, we obtain fidelities of $0.82(1)$
and $0.74(1)$ (without correcting for SPAM errors) for the unitary and adaptive state preparation circuits, respectively.
By comparison, a GHZ fidelity $>0.5$ is sufficient to witnesses genuine multipartite entanglement~\cite{Guehne2010}.
The adaptive circuit contains more 2Q gates (46) and measurements (48), and therefore produces a lower fidelity than the unitary circuit, which contains 31 2Q gates and 32 measurements. Especially for systems with limited connectivity and appreciable memory errors, the constant-depth adaptive circuit should outperform the unitary preparation circuit at large enough $N$.

\begin{table}[]
\caption{GHZ state fidelities.}
\begin{ruledtabular}
\begin{tabular}{lc}
GHZ prep &  State fidelity \\ \hline
$N=20$ & 0.86(1)             \\
$N=26$ & 0.83(1)             \\
$N=32$ & 0.82(1)             \\
$N=32$ adaptive & 0.74(1)              \\
\end{tabular}
\end{ruledtabular}

\label{GHZ_table}
\end{table}

\section{Application benchmarks} \label{sec:algobench}

The system-level benchmarks of the previous section
serve to verify quantum computer performance on a well-defined set of volumetric circuits.
However, problems of practical interest tend to involve structured circuits with very specific demands on gate set and connectivity. A comprehensive survey of all such problems is beyond the scope of this work (and difficult to define), but a sampling of such applications is still helpful for evaluating the machine's capabilities with respect to plausible near-term use cases and the demands they impose. In this section, we present the results of four application benchmarks: (A)
Hamiltonian simulation, (B) QAOA, (C) large-distance repetition codes, and (D) holographic quantum dynamics simulation.  We chose these benchmarks in a complementary way, as each places a different emphasis on particular error sources.  For example, Hamiltonian simulation is highly dependent on 2Q gate error, QAOA performance depends strongly on qubit connectivity, and repetition codes and the holographic quantum dynamics simulation require high-fidelity MCMR.

\subsection{Hamiltonian simulation}
Simulating the continuous time evolution of many-body quantum systems is an important and classically challenging problem for which quantum computers are well-suited \cite{Feynman82,Wecker15,Childs18}. To benchmark the performance of the H2 quantum computer on this task, we simulate the dynamics of an $L=32$ site transverse-field Ising model (TFIM)
in one spatial dimension, with Hamiltonian
\begin{align}
H=-J\sum_{j=1}^{L}Z_jZ_{j+1} - h \sum_{j=1}^L X_j.
\end{align}
Here and elsewhere in this section, site subscripts are taken ${\rm mod}(L)$ to yield periodic boundary conditions.
We simulate a quantum quench where the initial state is prepared in the ground state at $h/J=\infty$, that is, $\ket{\Psi(0)}=\ket{+}^{\otimes L}_j$.  The Hamiltonian is suddenly quenched to $h/J=0.2$, and the state is then evolved up to $Jt=7$ under the new Hamiltonian. We evaluate the dynamics of the expectation value of the Pauli $X$ operator averaged over all qubits, i.e.,
$\<X\>\equiv \frac{1}{L}\sum_{j=1}^{L} \<X_j\>$. We digitally simulate the dynamics using 1st-order Trotterization of the time-evolution operator~\cite{seth96}, 
\begin{align}
U(t)\approx\left(\prod_j \exp\left[i Z_jZ_{j+1} \frac{Jt}{ r}\right]\prod_j\exp\left[i X_j \frac{h t}{ r}\right]\right)^r,
\label{eq:trotter_formula}
\end{align}
which approaches the true evolution as $r\to \infty$. 
The 1Q and 2Q gates in this decomposition are
$X$ rotations and $U_{ZZ}(\theta)$,
which are native on H2.
The number of Trotter steps $r$ is chosen such that the errors on $\langle X \rangle$ due to Trotterization are below 0.01, as determined by explicit calculations of the noiseless Trotterized dynamics \cite{PhysRevA.65.032325} and comparison to exact results for the continuous-time evolutions \cite{Calabrese_2012}. This threshold ensures that Trotter errors are at or below the scale of the expected $\sim1\%$ statistical fluctuation in the experiment (more details in App.~\ref{step_err_detail}). 

\begin{figure}
\begin{center}
\includegraphics[width=0.45\textwidth]{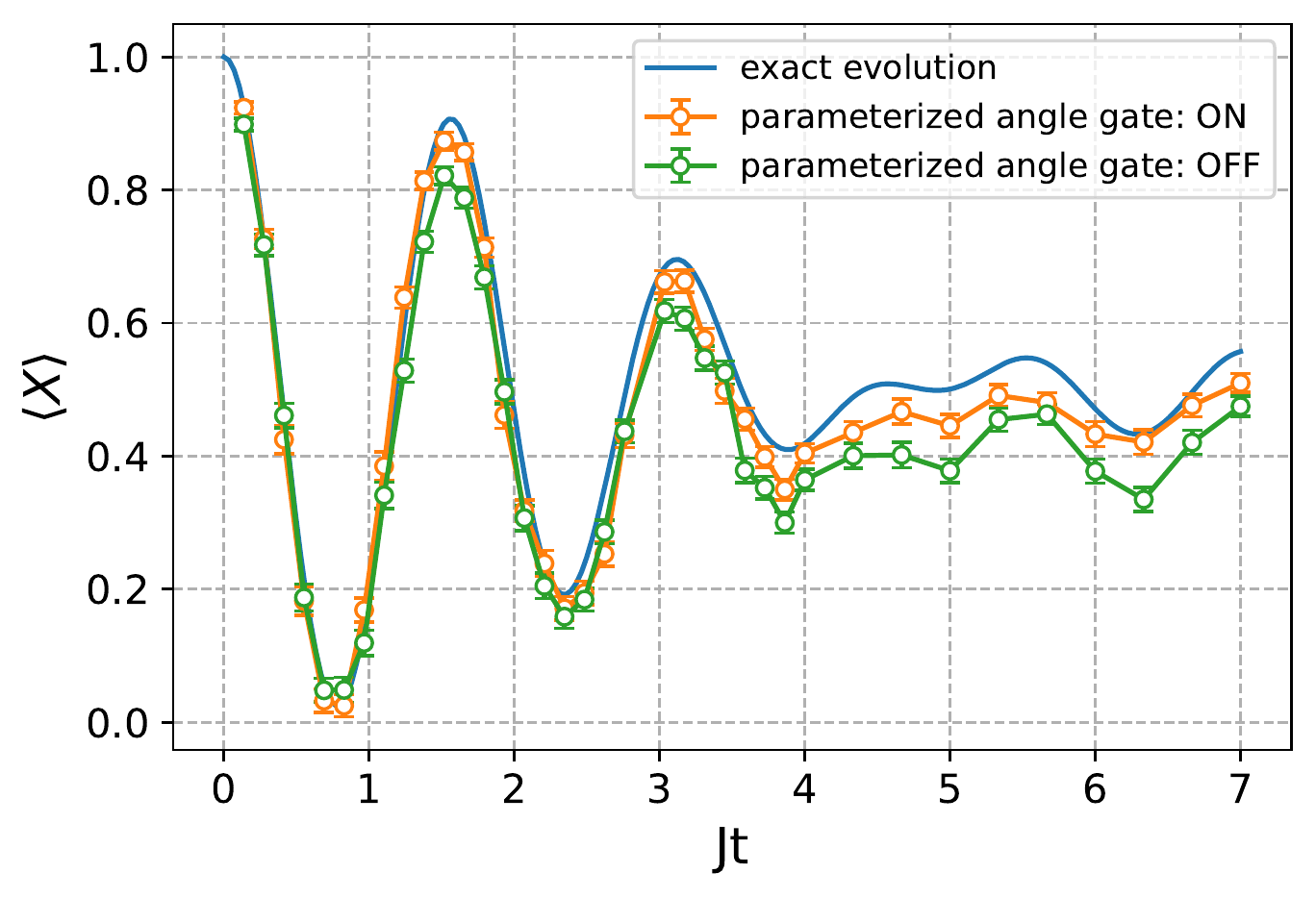}
\end{center}
\caption{The dynamics of $\<X\>$ for a 32-qubit TFIM Hamiltonian simulation vs.~evolution time. The orange data is obtained by directly implementing each $ZZ$ rotation in every Trotter step using our native parameterized angle $U_{ZZ}(\theta)$ gate with $\theta=2Jt/r$. The green data is obtained by decomposing each $ZZ$ rotation into two $U_{ZZ}(\pi/2)$ (Clifford) gates with some 1Q rotations. Each data point is obtained as the average of 100 shots of the associated Trotterized circuit for that time.} 
\label{fig:TFIM}
\end{figure}
The results of our experiment, plotted in  Fig.~\ref{fig:TFIM}, show reasonably good agreement between our quantum simulation and the exact solution up to time $Jt=7$, suggesting the quantum computer has small enough errors to coherently simulate quantum dynamics up to a non-trivial time (note that a completely depolarized state has $\<X\>=0$). The data has not been post-processed or error-mitigated in any way. Figure \ref{fig:TFIM} also compares the circuit implementations with and without using the parameterized angle $U_{ZZ}(\theta)$ gate. Without parameterized angle gates, every such gate has to be decomposed into two $U_{ZZ}(\pi/2)$ gates with additional 1Q rotations, resulting in a doubling of the number of 2Q gates, and the possibility of more than doubling the error per Trotterization step (see Fig.~\ref{fig: arbrb decay}). The improvements to the simulation results when using parameterized-angle 2Q gates highlights their benefit for near-term applications of quantum computers to simulating many-body physics.

\subsection{QAOA}

The quantum approximate optimization algorithm (QAOA) \cite{original_qaoa} is a near-term heuristic algorithm for solving combinatorial optimization problems of general interest in many industries. As in previous benchmarking studies \cite{google1, google2, DeCross2022}, we focus on solving the MaxCut problem restricted to the class of unweighted 3-regular graphs $G = (V,E)$. The standard QAOA circuit consists of alternating applications of a mixing unitary $U_B(\beta_n) = e^{-i\beta_n H_B}$ and a phase-splitting cost unitary $U_C (\gamma_n) = e^{-i\gamma_n H_C}$,
\begin{align}
    U(\vec{\beta}, \vec{\gamma}) = \prod_{n=1}^p U_B(\beta_n) U_C (\gamma_n).
\end{align}
The $2p$ parameters $\beta_n$ and $\gamma_n$ are found variationally, by searching with a classical optimization algorithm for the choice of parameters that extremizes the cost of the QAOA final state,
\begin{align}
    \langle H_C \rangle = \langle \psi_0 | U(\vec{\beta}, \vec{\gamma})^{\dagger} H_C U(\vec{\beta}, \vec{\gamma}) | \psi_0\rangle.
\end{align}
The initial state is taken to be $|\psi_0\rangle = |+\rangle^{\otimes N}$, the ground state of $H_B = \sum_i X_i$, while for the unweighted MaxCut problem the cost Hamiltonian is
\begin{align}
    H_C = \frac12 \sum_{(i,j) \in E}  (1-Z_i Z_j),
\end{align}
and therefore each term in the cost Hamiltonian comprising the cost unitary $U_C(\gamma_n)$ can be implemented with a single $U_{ZZ}(\theta)$ gate.

For the classical optimization procedure, we use the derivative-free BOBYQA optimizer \cite{BOBYQA} as implemented in the \verb|Py-BOBYQA| package \cite{PyBOBYQA}. The BOBYQA optimizer builds a local quadratic model to the objective function within a trust region of size that decreases with iterations of the optimizer. We set the optimizer convergence conditions to be met when the precision of the variational parameters reaches the same order as the measured 2Q gate errors, $1 \times 10^{-3}$.

We study two separate experiments in this work. The first implements a larger-scale MaxCut QAOA problem ($N = 130$, $p = 1$) on 32 physical qubits using qubit-reuse compilation \cite{DeCross2022} and 100 shots per circuit. The second experiment solves an $N = 32$ MaxCut QAOA problem at $p = 2$ with 200 shots per circuit to demonstrate an improvement in solution quality compared to $p = 1$ with the more expressive and deeper ansatz. For plotting purposes, the energy was rescaled by a sign so that in all cases the optimum corresponds to the solution of minimum energy.

\begin{figure}[]
\begin{center}
\includegraphics[width=0.45\textwidth]{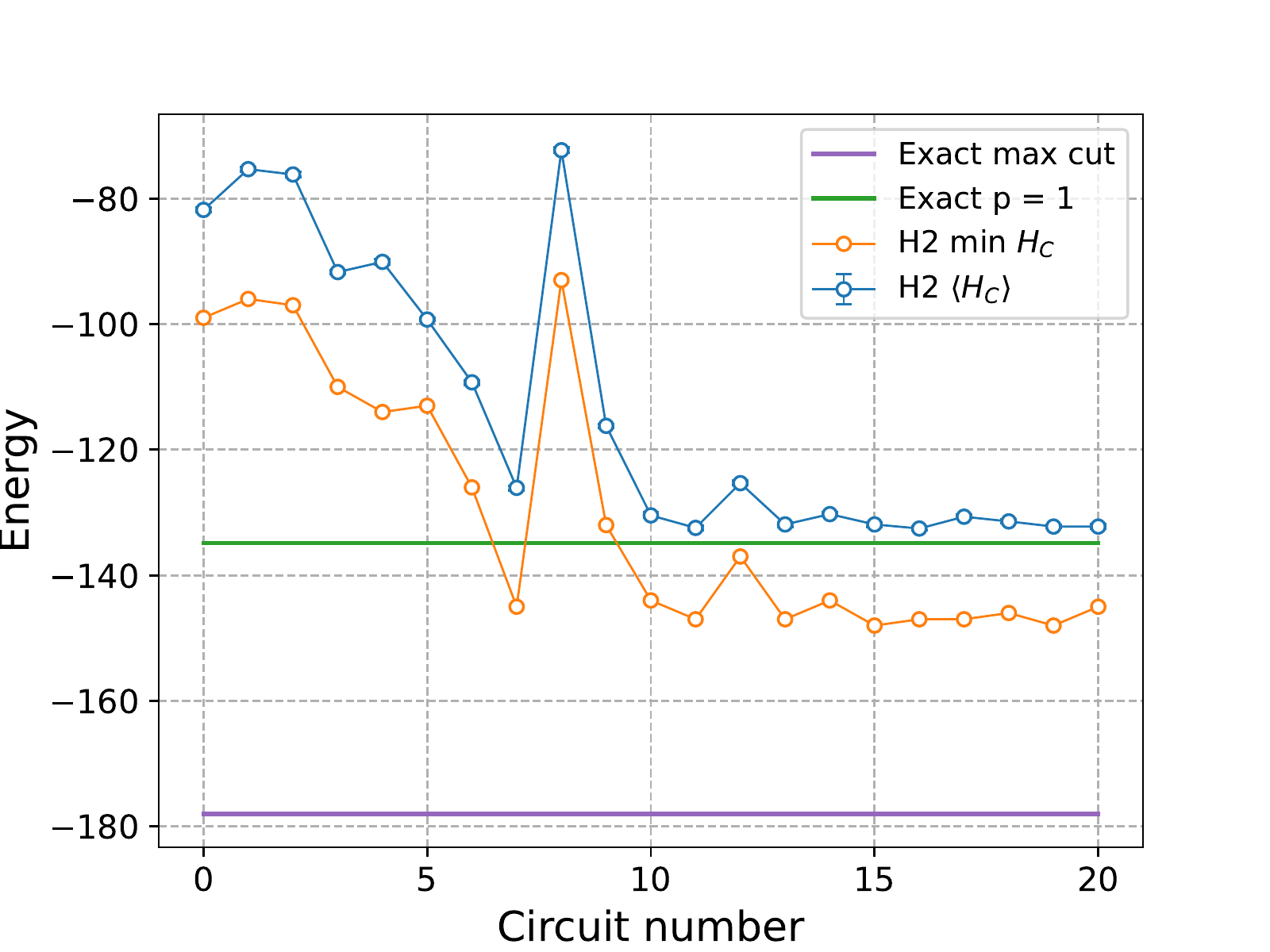}
\end{center}
\caption{Optimization trajectory of $N = 130$, $p = 1$ QAOA computed via qubit reuse on H2. The expectation value of the energy as measured experimentally at $p = 1$ (blue) converges well to the best possible exact value (green). Uncertainties on the measured value of $\langle H_C \rangle$ are plotted but smaller than the displayed point size (see App.~\ref{app:qaoa} for details). The best sample taken at each iteration (orange) is also displayed relative to the true max cut (purple).}
\label{fig:N130QAOA}
\end{figure}

\begin{figure}[h!]
\begin{center}
\includegraphics[width=0.45\textwidth]{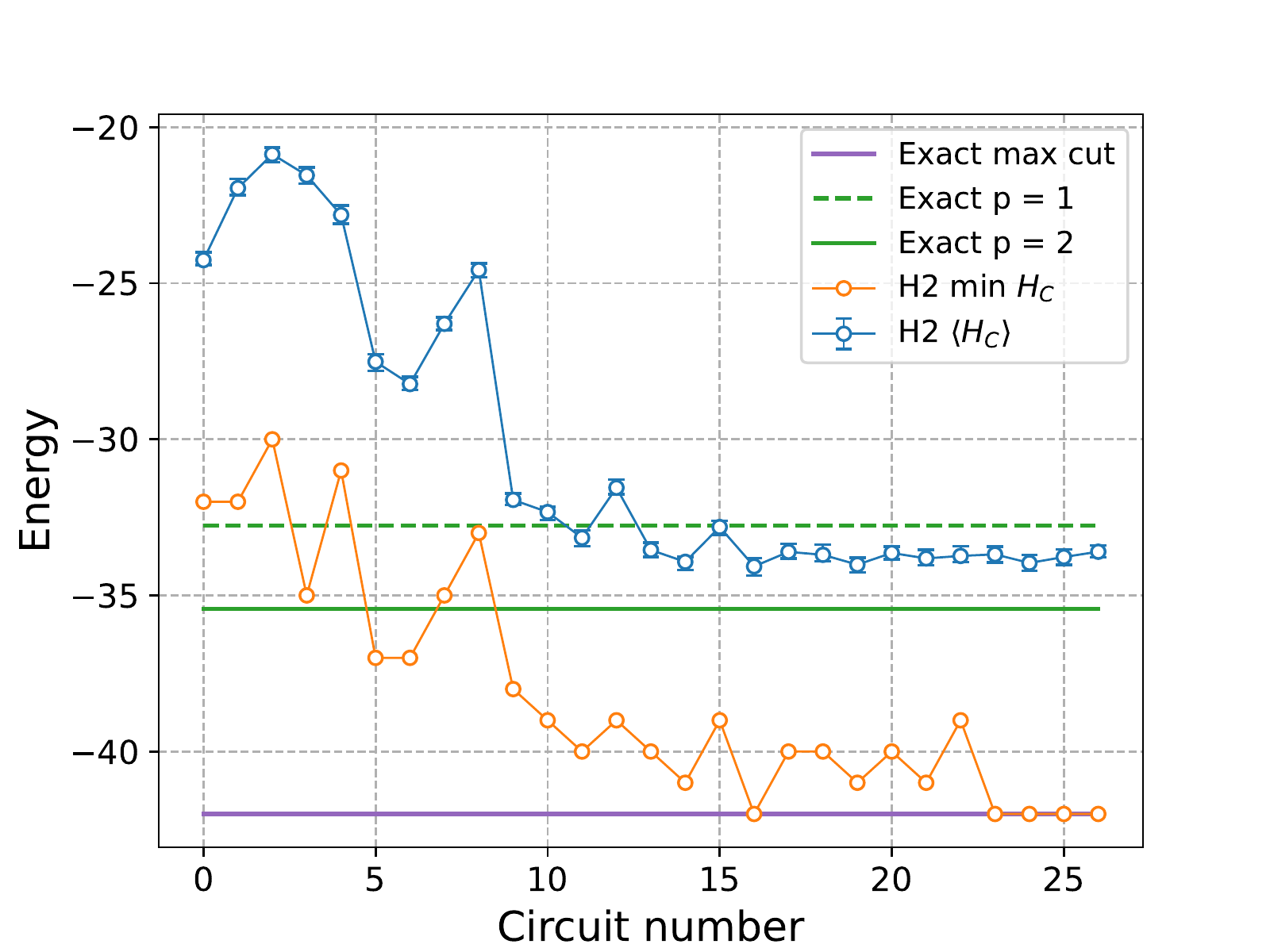}
\end{center}
\caption{Optimization trajectory of $N = 32$, $p = 2$ QAOA on H2. The expectation value of the energy as measured experimentally at $p = 2$ (blue) surpasses the best possible exact value for a $p = 1$ circuit (green, dashed). The best sample taken at each iteration (orange) is also displayed relative to the true max cut (purple).}
\label{fig:N32QAOA}
\end{figure}

In Fig.~\ref{fig:N130QAOA} we display the results from the $N = 130$, $p = 1$ experiment. The optimizer shows  convergence within the first ten circuits. Using the tensor network methods available in the Python library \verb|quimb|~\cite{gray2018quimb} in conjunction with the global Bayesian optimizer in \verb|scikit-optimize| \cite{tim_head_2018_1207017} we also exactly evaluated the best average energy possible for any $p = 1$ circuit. The convergence of the blue data to the green line in Fig.\,\ref{fig:N130QAOA} demonstrates that the optimization procedure succeeded in locating the optimal parameters and that H2 evaluated the circuits with sufficiently low noise to nearly saturate the best possible result. In App.~\ref{app:qaoa} we also display the optimization trace on the energy landscape, further confirming that the optimizer succeeded in locating the optimal parameters. To evaluate the performance of the algorithm in solving the combinatorial problem, we also compare the minimum value of the energy sampled in any given shot to the exact value of the max cut computed in \verb|gurobi| \cite{gurobi}. As expected since the circuit depth is only $p = 1$, the best cut value found on H2, 148, is substantially less than the exact value of 178. Nevertheless, this experiment represents substantial progress towards solving industry-scale combinatorial problems with QAOA on small quantum computers.

In Fig.~\ref{fig:N32QAOA} we demonstrate the results of the $N = 32$, $p = 2$ optimization procedure. Comparing to the best average energies possible for any $p = 1$ or $p = 2$ circuits, the experimental data for $p = 2$ consistently performs better than the best possible $p = 1$ circuit and is close to saturating the ground state energy for $p = 2$ circuits. Furthermore, H2 succeeded in locating solutions with the best possible max cut of 42 for this graph.

\subsection{Error correction: repetition code} 
Large quantum computations are widely thought to only be possible through quantum error correction (QEC).
Therefore, in the context of fault-tolerant quantum computers, perhaps the most important quantum algorithm is not a particular targeted calculation, but rather the QEC algorithm being run in the background. Additionally, given the large resource overheads of QEC, the design requirements for large scale quantum computers will likely be driven by the optimization of these codes' power and efficiency, highlighting the importance of closely related benchmarks.

\begin{figure}[]
\begin{center}
\includegraphics[width=0.9\columnwidth]{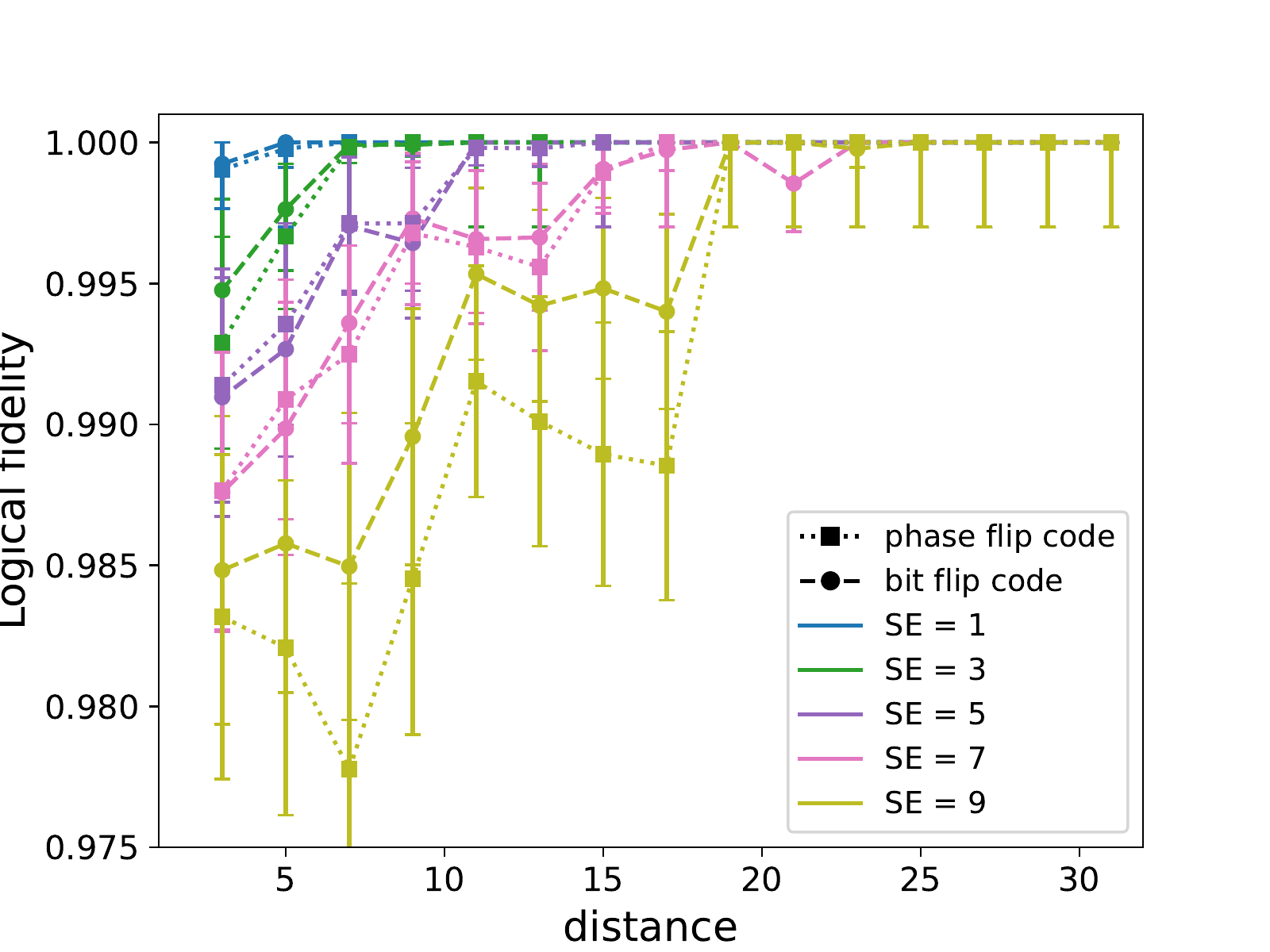}
\end{center}
\caption{The logical fidelities of the phase (square, dotted) and bit (circle, dashed) flip repetition code as a function of distance. As the number of syndrome extraction (SE) rounds increases, more noise is injected into the system, degrading the logical fidelity. For a given number of rounds of SE, as the code distance increases, so does the logical fidelity. All error bars are calculated using Jack-knife resampling \cite{efron1982jackknife}, except for those where the sampling number was too low to calculate an error (i.e. 100$\%$ fidelities), in which case the statistical rule of three was used \cite{eypasch1995probability}.}
\label{fig:RepLogFvsD}
\end{figure}

Many QEC schemes are based on stabilizer codes that encode logical information into the joint subspace of many physical qubits, known as data qubits. Additional physical qubits, known as ancilla qubits, are used to make non-destructive syndrome measurements~\cite{gottesman1997stabilizer} which discretize errors into a manageable set of bit and phase errors, allowing for general QEC.
Repetition codes are examples of stabilizer codes but can only correct a single type of error, typically either bit or phase flip errors.
However, they make good benchmark algorithms since they possess all the components needed to implement a quantum code. Specifically, a distance $d$ repetition code can reliably correct up to $\lfloor \frac{d-1}{2}\rfloor$ errors.  Corrections are determined by repeatedly measuring stabilizers of the code using MCMR, syndromes are decoded using algorithms similar to those used in quantum codes, and calculating logical fidelities is done in the usual way. Using all 32 qubits, we implement a $d=31$ repetition code with 31 data qubits and one ancilla, maximizing the code distance that can be tested. This low overhead implementation of the code is made possible by H2's qubit reuse capabilities and performing 30 unique stabilizer measurements serially.

\begin{figure*}
\begin{center}
\includegraphics[width=\textwidth]{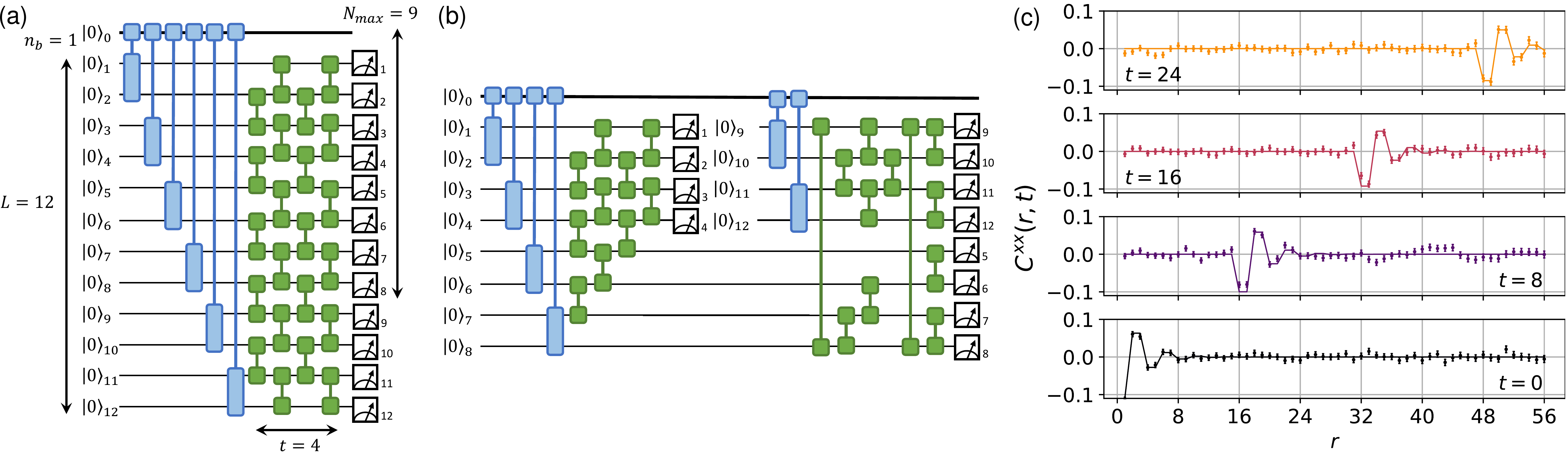}
\end{center}
\caption{(a) A one-dimensional brickwork circuit of length $L=12$ with $t=4$ layers of gates applied to a quantum matrix product state of bond-dimension $\chi=2^{n_b}=2$. (b) HoloQUADS re-uses qubits through MCMR to execute the same circuit with a minimal number of qubits. Here we use $N_{max}=9$ qubits, but $N_{max}$ can be adjusted between $n_b+t+2=7$ (maximally serial) and $n_b+L=13$ (maximally parallel). (c) The experimentally measured (dots) correlation function $C^{xx}(r,t)$ for a dual-unitary circuit applied to a length $L=128+t$ solvable $\chi=2$ quantum matrix product state compared to the exact thermodynamic limit results (solid lines), using $N_{max}=32$ qubits up to time $t=24$. Error bars are standard deviations of the mean from four 100 shot experiments.}
\label{fig:holoQUADS}
\end{figure*}

The syndrome measurements are processed in real-time using Wasm calls to the classical compute environment during the quantum circuit. At the end of the circuit, the data qubits are also measured and used to construct a final syndrome measurement.
We use this last syndrome in addition to all previously recorded syndromes to decode the logical output state and calculate the logical fidelities. The decoding uses a minimum-weight perfect-matching algorithm \cite{dennis2002topological, higgott2022pymatching}, which is performed online at the end of the circuit as part of the control-system software execution of each shot (i.e. while the hybrid quantum/classical program is still being executed on the actual device). Since this is done after the logical qubit has been measured, the operation does not perform mid-circuit real-time decoding, making these experiments insensitive to memory error associated with the computation time of a correction. Real-time decoding operations are possible with Wasm and the advanced classical compute environment infrastructure, but they are unnecessary for repetition code memory experiments. 

Experiments on both the $d=31$ bit flip code and phase flip code were performed while varying the number of rounds of syndrome extraction, and recording all syndrome measurements, allowing us to process subsets of the code after the program completes. The subsets allow us to reconstruct logical fidelities for all odd distance codes less than $d = 31$. These measurements are similar to Ref.~\cite{chen2021exponential, Acharya2023}, which use a fixed architecture and parallel syndrome measurements, making for a direct comparison of different distances. In contrast, our architecture offers a less direct comparison between different code distances, as syndrome measurements are done serially, but allows for larger distance codes with lower qubit overheads. We note that the Wasm decoder was only used to calculate the $d=31$ fidelities. All other code distance fidelities were calculated by the same minimum-weight perfect-matching algorithm offline. 

The experimental results in Fig.~\ref{fig:RepLogFvsD} show the larger distance codes achieve higher logical fidelities as expected, with the bit flip code producing a higher logical fidelity compared to the phase flip code for a given distance, consistent with a biased noise environment. These results demonstrate many of the necessary components for implementing scalable, real-time QEC, and show how the capabilities of the H2 system can help realize large distance stabilizer QEC codes, all of which will be the subject of future studies.

\subsection{Holographic quantum dynamics simulation (HoloQUADS)}

High fidelity MCMR is crucial for quantum error correction, and can also help expand the reach of many near-term algorithms \cite{DeCross2022}. In particular, such techniques have been shown to enable the simulation of quantum dynamics from initially correlated states directly in the thermodynamic limit, with qubit number requirements set by the evolving entanglement entropy of the state rather than its physical size \cite{FossFeig2021}. Based on work in Refs. \cite{bertini2018,Bertini2019}, Ref. \cite{HoloChertkov2022} recently proposed and demonstrated a benchmark for such methods by simulating exactly solvable dual-unitary circuit models applied to initial matrix-product states on H1-1.  Here we use the additional resources of H2 to extend those results to longer evolution times, where the system contains more entanglement.

Following Ref.~\cite{HoloChertkov2022}, we simulate time evolution under dual-unitary circuits~\cite{bertini2018,Bertini2019} which are one-dimensional brick-work circuits (Fig.~\ref{fig:holoQUADS}a) having generic properties of typical circuits (e.g., exhibiting quantum chaos and ballistic growth of entanglement ~\cite{bertini2018}) and certain non-generic properties (e.g., their correlations spread at the maximal possible velocity \cite{claeys2020}, and are confined to the light-cone boundary rather than its interior), which allow quantities such as entanglement entropy and correlation functions to be analytically determined \cite{bertini2018,Bertini2019,Gopalakrishnan2019}. An initial matrix product state is prepared by applying gates between the physical qubits and an ancilla ``bond'' qubit (blue gates in Fig.~\ref{fig:holoQUADS}a,b) and then time-evolving this state by the self-dual kicked Ising (SDKI) model \cite{Akila2016,bertini2018} (green gates in Fig.~\ref{fig:holoQUADS}a,b). After $t$ layers of SDKI gates are applied to $\ket{\psi_0}$, we measure the smoothed correlation functions
\begin{equation}
C^{xx}(r,t) = \frac{1}{2L}\sum_{j=1}^{L} \sum_{\delta=0,1} \langle\psi_t|X_{j}X_{j+r+\delta}|\psi_t\rangle, \label{eq:Cab0}
\end{equation}
 where $|\psi_t\rangle$ is the time-evolved state and $L$ is the system size, and in Fig.~\ref{fig:holoQUADS}c we compare the results to exact theoretical calculations from Ref.~\cite{Bertini2019}. We use H2's 32 qubits to simulate $t=0,8,16,24$ layers of time-evolution applied to a length $L=128+t=128,136,144,152$ matrix product state.

The experimental data show close agreement with ideal noiseless results, suggesting H2's mid-circuit measurement, mid-circuit reset, crosstalk, and memory errors are low enough for sizeable quantum dynamics simulations using HoloQUADS. We note that effects of errors can be highly circuit dependent. For the particular dual-unitary circuit studied in this benchmark, their maximal velocity behavior \cite{claeys2020} causes only $\propto t$ Pauli errors along the edges of the causal cones of qubits $i$ and $j$ to affect the $\langle \psi_t| X_i X_j |\psi_t\rangle$ correlation function. For a generic circuit, we would expect any of the $\propto t^2$ Pauli errors in the causal cones to affect correlation functions, meaning dual-unitary circuits are less sensitive to errors than typical circuits.

\section{A Summary of the Results and our Outlook}

The H2 quantum computer is a significant upgrade from our previous H1 system, maintaining or exceeding many previous fidelity metrics while operating on more qubits.  The clearest manifestation of the robust scaling of our QCCD architecture is that the system-level benchmarks are consistent with the errors measured by the component benchmarks. We also benchmarked H2's performance on a variety of applications that are widely considered to be well-suited for near-term quantum computers, with the goal of assessing the feasibility of such algorithms given current hardware performance metrics. Our benchmarking results show that 2Q gates remain the dominant error source, although fidelities improved slightly in our new generation. However, the transport time between arbitrary circuit layers did increase, which translates to larger memory error. Future work will focus on reducing both of these error sources by improvements to laser systems, transport speed, and magnetic field stability.

The new system also demonstrates a number of key technological milestones on the path to scaling, including ion transport controlled via broadcast electrode signals, RF signals routed under the surface of the trap, and fast MOT-based loading. These improvements are achieved in a system initially configured to operate with 32 qubits (but designed to accommodate more) and collectively bolster the case for the viability of the QCCD architecture as a route to large-scale trapped-ion quantum computing. The further development of the QCCD architecture will include truly two-dimensional trapping structures for fast ion sorting~\cite{burton2022}, as well as moving beyond free-space optical delivery.\\[.2cm]

\section*{Data availability}
All data presented is available in Ref.~\cite{github_h2}. Most component benchmarking data is available in Ref.~\cite{github_spec} and will be updated as H2 improvements are introduced. Quantum volume data is available in Ref.~\cite{github_qv} and will be updated as new tests are run.

\begin{acknowledgments}
We thank the entire Quantinuum team for numerous contributions that enabled this work, and we thank Matt Marcus, Philip Makotyn, and Tom Loftus for helpful and inspiring conversations.  We thank Jack Ross for creating Fig.~\ref{fig:prettytrappic} and Eric Hudson for helpful comments on an earlier version of the manuscript.
\end{acknowledgments}

\appendix

\section{Details of component benchmarks}

\begin{table*}[]
\begin{ruledtabular}
\begin{tabular}{lccc}
Test                  & Lengths                          & ~Repetitions~ & Shots/Circuit \\ \hline
1Q RB                 & {[}2, 32, 128, 512{]}            & 30          & 100           \\
2Q RB                 & {[}2, 16, 64, 128{]}             & 30          & 100           \\
2Q SU(4) RB           & {[}2, 8, 32, 64{]}               & 15          & 100           \\
2Q parameterized RB & {[}4, 50, 100{]}                 & 10          & 100           \\
Transport 1Q  RB       & {[}2, 16, 32, 64{]}              & 64*         & 200           \\
Measurement crosstalk & ~{[}0, 100, 200, 300, 400, 500{]}~ & 1           & 1000          \\
Reset crosstalk       & ~{[}0, 100, 200, 300, 400, 500{]}~ & 1           & 1000          \\ 
SPAM     & - & 2           & 5000          \\ 
\end{tabular}
\end{ruledtabular}
\caption{\label{tab:rb_specs}Parameters used for component benchmarking testing. *The transport 1Q  RB test is done with 32 qubits in parallel and repeated twice.}
\end{table*}

\subsection{Randomized benchmarking parameters and data} \label{app:RB_data}
All component benchmarks (except transport 1Q  RB) were repeated for each gate zone (DG01-DG04). Transport 1Q  RB used all available 32 qubits with random rearrangements so is not zone specific. For each RB experiment, sequences were randomly and independently generated for each qubit (or pair of qubits for all tests with 2Q gates).  The sequence lengths, repetitions, and shots used for the component benchmarks are shown in Table~\ref{tab:rb_specs}. 

For each RB experiment the decay is fit to the standard first order RB function~\cite{Magesan11} with a fixed asymptote,
\begin{equation}\label{eq: generic RB fit}
    p(\ell) = Ar^{\ell} + 1/2^N,
\end{equation}
where $p(\ell)$ is the observed survival probability at length $\ell$, $A$ is the SPAM fit parameter, $r$ is the depolarizing rate and $N$ is the number of qubits. The reported error is the average infidelity, which is given by
\begin{equation}\label{eq: RB average fidelity}
    \epsilon = \frac{2^N - 1}{2^N}(1 - r).
\end{equation}
This captures only the errors in the computational subspace and not leakage errors, which are measured with the leakage detection gadget.

The reset and measurement crosstalk decay functions are fit to functions derived from error models of their respective operations in Ref.~\cite{Gaebler2021} with the following equations
\begin{align}
    p_M(\ell) &= \tfrac{1}{3}(2 - A_M + (4A_M - 2)e^{-3 r_M\ell}) , \\
    p_R(\ell) &= 1 - A_R + \tfrac{1}{3}e^{-5 r_R \ell}(2 + e^{3r_R\ell})(2 A_R - 1),
\end{align}
where $A_{M/R}$ are the SPAM fit parameters for each method and $r_{M/R}$ is the rate of measurement/reset crosstalk scattering. Each scattering rate is then converted to average infidelity
\begin{align}
    \epsilon_M &= 5 r_M/6, \\
    \epsilon_r &= 5 r_R/3.
\end{align}

\begin{figure}
\includegraphics[width=0.9\columnwidth]{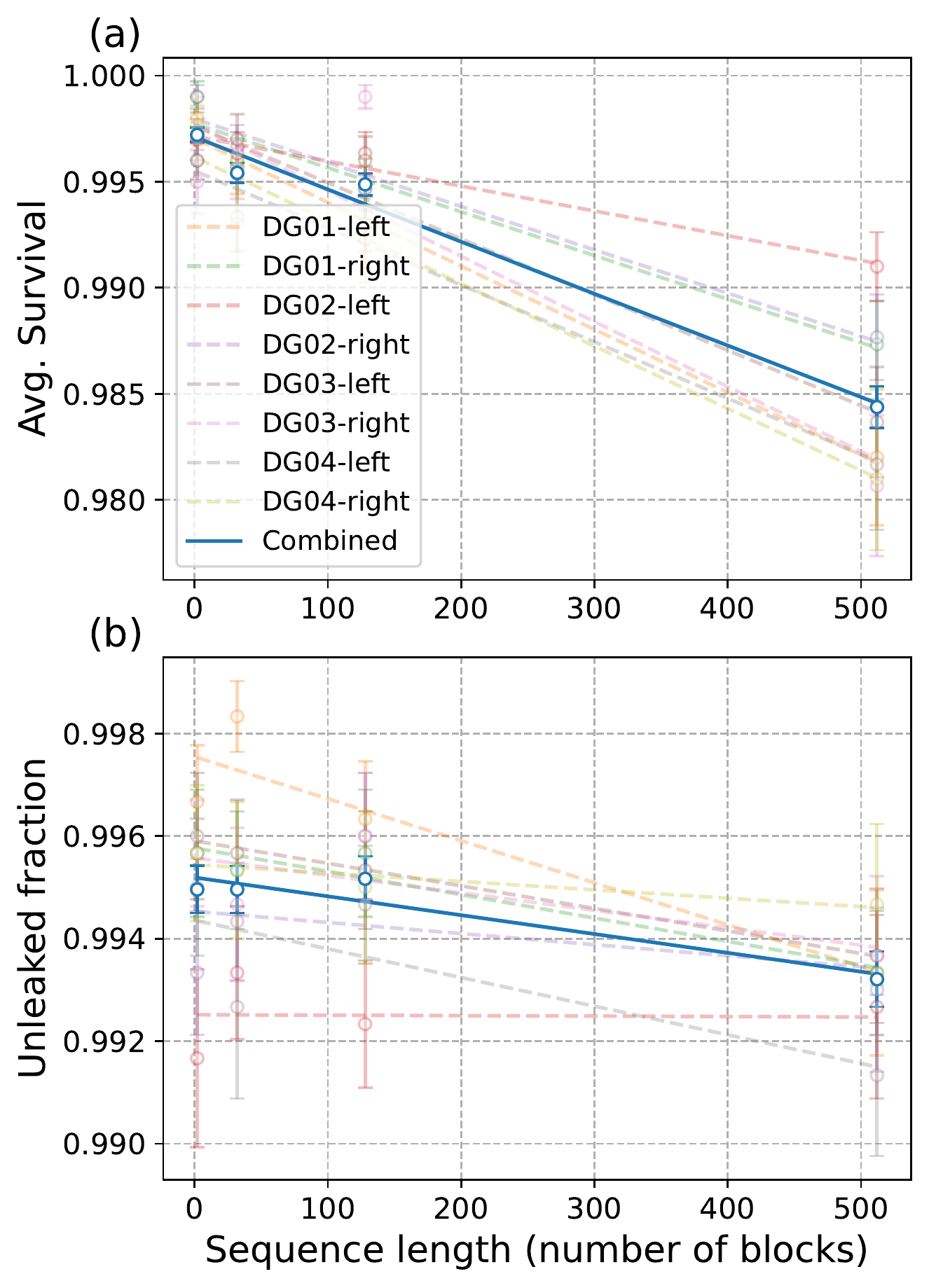}
\caption{1Q RB data with parameters given in Table~\ref{tab:rb_specs}. (a) Decay of survival probability. (b) Decay of unleaked fraction of shots.}
\label{fig:1Q rb}
\end{figure}

\begin{figure}
\includegraphics[width=0.9\columnwidth]{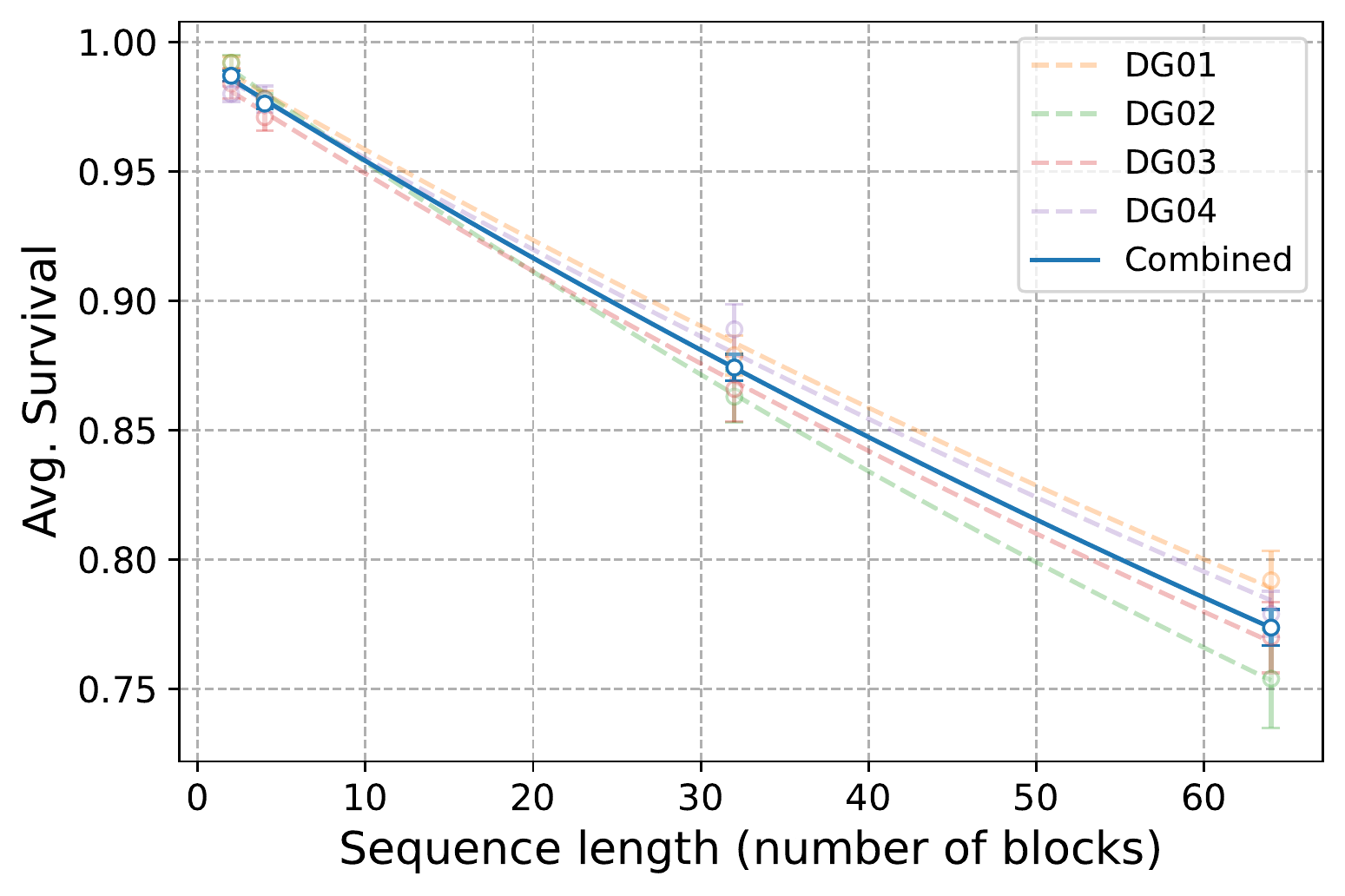}
\caption{2Q SU(4) RB data with parameters given in Table~\ref{tab:rb_specs}.}
\label{fig:su4 rb}
\end{figure}

\begin{figure}
\includegraphics[width=0.9\columnwidth]{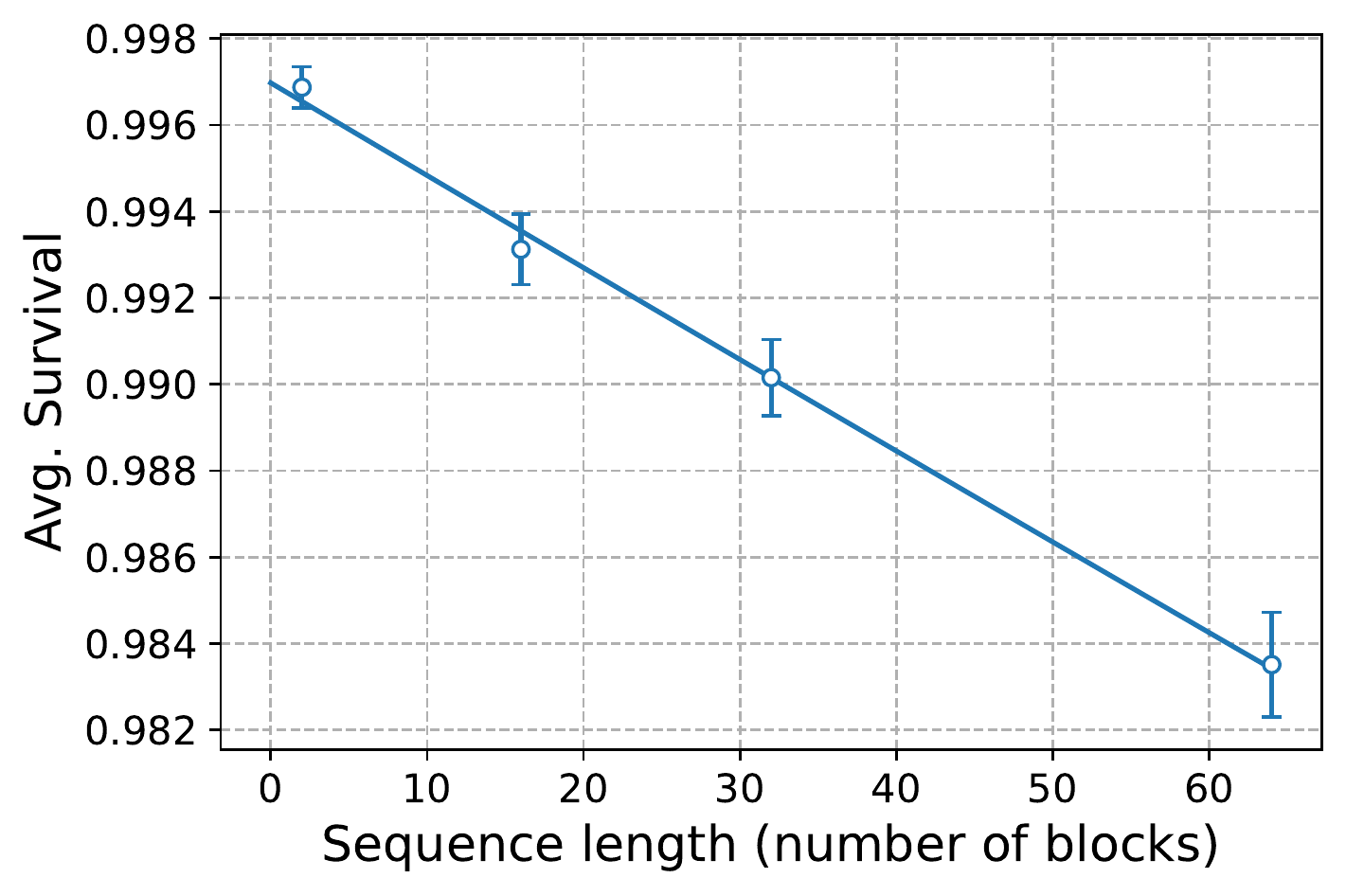}
\caption{Transport 1Q RB with parameters given in Table~\ref{tab:rb_specs}.}
\label{fig:transport 1Q rb}
\end{figure}

\begin{figure}
\includegraphics[width=0.9\columnwidth]{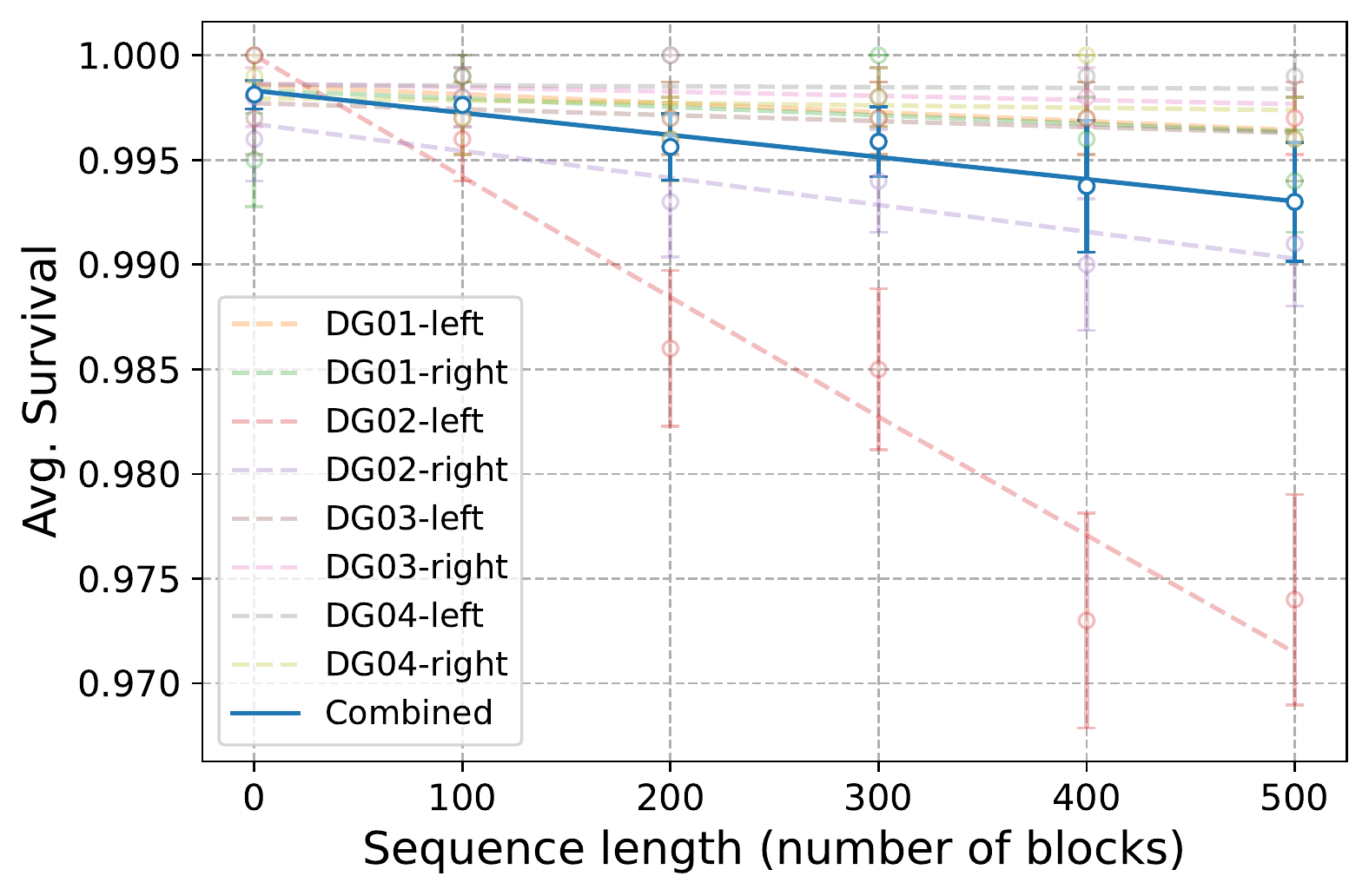}
\caption{Measurement crosstalk data with parameters given in Table~\ref{tab:rb_specs}.}
\label{fig:measurement crosstalk}
\end{figure}

\begin{figure}[]
\includegraphics[width=0.9\columnwidth]{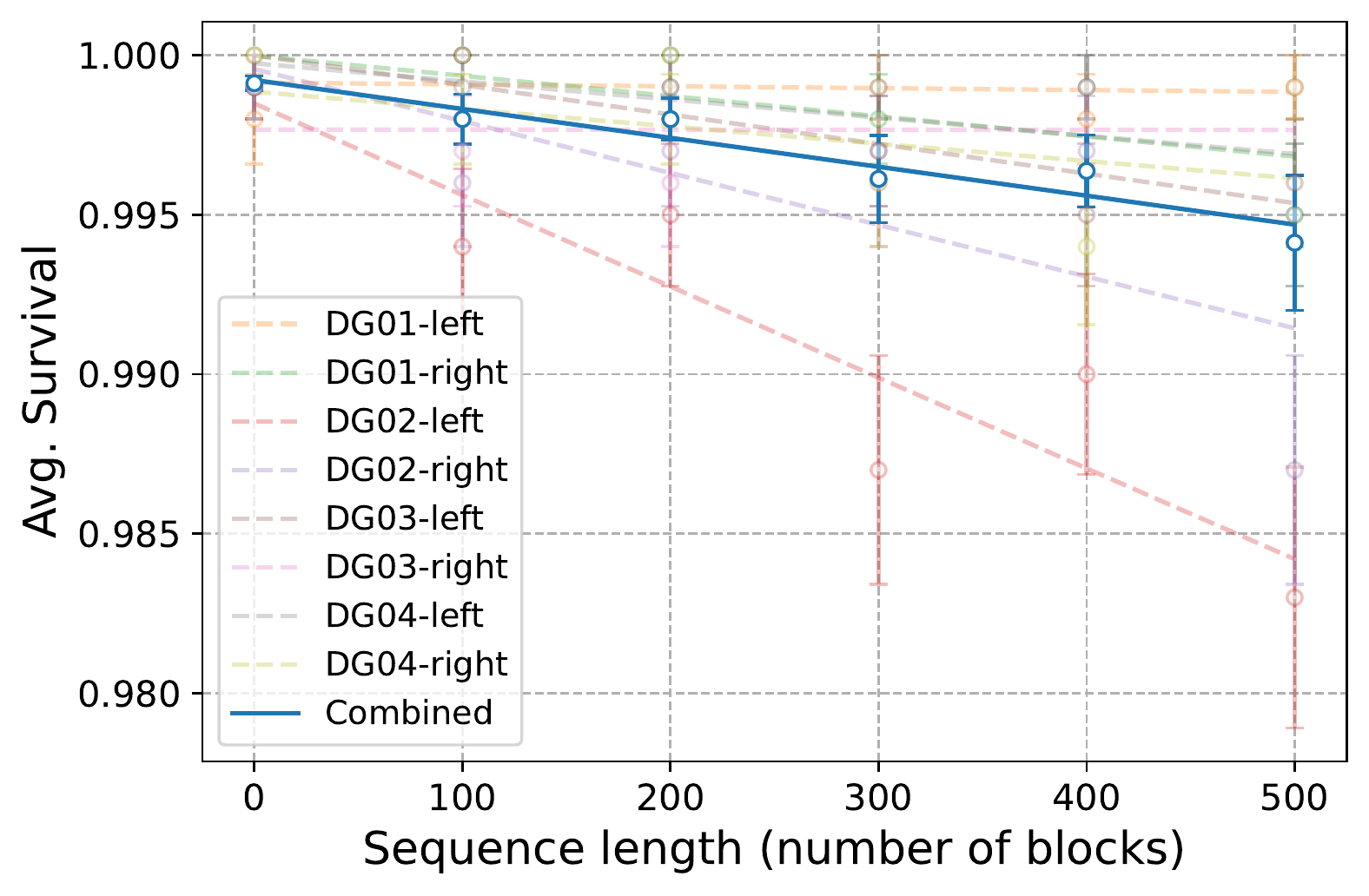}
\caption{Reset crosstalk data with parameters given in Table~\ref{tab:rb_specs}.}
\label{fig:reset crosstalk}
\end{figure}

For all component measurements a final combined estimate is obtained by performing the RB (or crosstalk) analysis on a combined dataset  between all measured qubits, which is reported in Table~\ref{tab:RB_avgs}. For example, in 1Q RB the combined dataset is obtained by treating each qubit measurement as a single sequence randomization and performing the RB fitting averaged over every qubit's random sequences. This leads to an RB experiment with $8\times 40$ random sequences for each length. For the crosstalk and SPAM measurements the combined dataset is obtained by adding all circuit output counts together. Zone specific data for each component testing experiment is shown in Table~\ref{tab:rb_full_data}. Decay plots for each component benchmark are shown in Figs.~\ref{TQ_RB_decay},~\ref{fig:1Q rb},~\ref{fig:su4 rb},~\ref{fig:transport 1Q rb},~\ref{fig:measurement crosstalk} and~\ref{fig:reset crosstalk}.

\begin{table*}[]
\begin{ruledtabular}
\begin{tabular}{lccccc}
 Test                   & DG01                     & DG02                      & DG03                    & DG04                      & Combined  \\ \hline
1Q  RB                    & [0.30(9), 0.21(7)]    & [0.12(5), 0.21(7)]     & [0.27(8), 0.32(9)]   & [0.27(9), 0.30(9)]     & 0.25(3)   \\
1Q  leakage rate               & [0.08(5), 0.04(4)]     & [0.00(4), 0.002(4)] & [0.04(5), 0.03(4)]  & [0.05(5), 0.02(4){]}       & 0.04(2)   \\
2Q RB                    & 19(1)                 & 17.5(9)                  & 18(1)                  & 17.9(9)                  & 18.3(5)   \\
2Q leakage rate               & 3.6(4)                   & 3.9(4)                    & 4.2(5)                  & 4.0(5)                    & 3.9(2)    \\
2Q SU(4) RB           & 39(3)                   & 45(3)                    & 42(3)                  & 38(2)                    & 41(1)    \\
Transport 1Q RB & -                       & -                        & -                      & -                        & 2.2(3)    \\
Measurement crosstalk    & [0.02(1), 0.02(2)]  & [0.24(3), 0.05(2)]   & [0.01(1), 0.01(1)] & [0.001(5), 0.005(9)] & 0.045(6) \\
Reset crosstalk          & [0.002(7), 0.02(1)] & [0.12(3), 0.007(2)]  & [0.04(1), 0.00(2)] & [0.02(1), 0.02(1){]}   & 0.038(6) \\
SPAM                     & [15(4), 16(4)]  & [19(4), 20(4)]   & [19(4), 16(3)] & [8(3), 12(3)]    & 16(1)  \\ 
\end{tabular}
\end{ruledtabular}
\caption{\label{tab:rb_full_data}Component benchmarking results for the tests outlined above. All values are in terms of average infidelity and $\times 10^{-4}$. For 1Q RB, 1Q leakage rate, measurement and reset crosstalk, and SPAM the brackets show the average infidelity for each side of the gate zone.}
\end{table*}

\subsection{Leakage detection gadget} \label{app:leakage}
The leakage rate $r_L$ is defined as the rate that population leaves the computational subspace due to a process $\Lambda$ based on Ref.~\cite{Wood18}
\begin{equation}
    r_L = \tfrac{1}{d_C}\textrm{Tr}(\mathds{1}_L \Lambda[\mathds{1_C}]),
\end{equation}
where $\mathds{1}_{L/C}$ is the identity operator on the leakage/computational subspace. The number of leakage detection events is fit to the model 
\begin{equation}
    p(\ell)=A(1-r_L)^\ell,
\end{equation}
as shown in Fig.~\ref{TQ_RB_decay}b and~\ref{fig:1Q rb}b. Gate errors in the leakage detection gadget can cause false-positive or false-negative detection events, but these only contribute to the parameter $A$, as they are independent of $\ell$, similar to the SPAM parameter in RB.

\subsection{2Q parameterized randomized benchmarking} \label{app:arb_rb}

To measure the average infidelity of $U_{ZZ}
(\theta)$ as a function of $\theta$,
we use direct RB. In standard RB, the 
unitaries comprising the RB sequence are
sampled from a unitary 2-design,
such as the Clifford group or SU($2^N$).
In contrast, direct RB samples unitaries from a set of native gates that generate the group~\cite{Knill2007, Proctor2019}.
Under such circumstances the survival probability
will still approximate an 
exponential decay with decay parameter linearly 
related to the average fidelity~\cite{polloreno2023}.

\begin{figure}
\includegraphics[width=0.9\columnwidth]{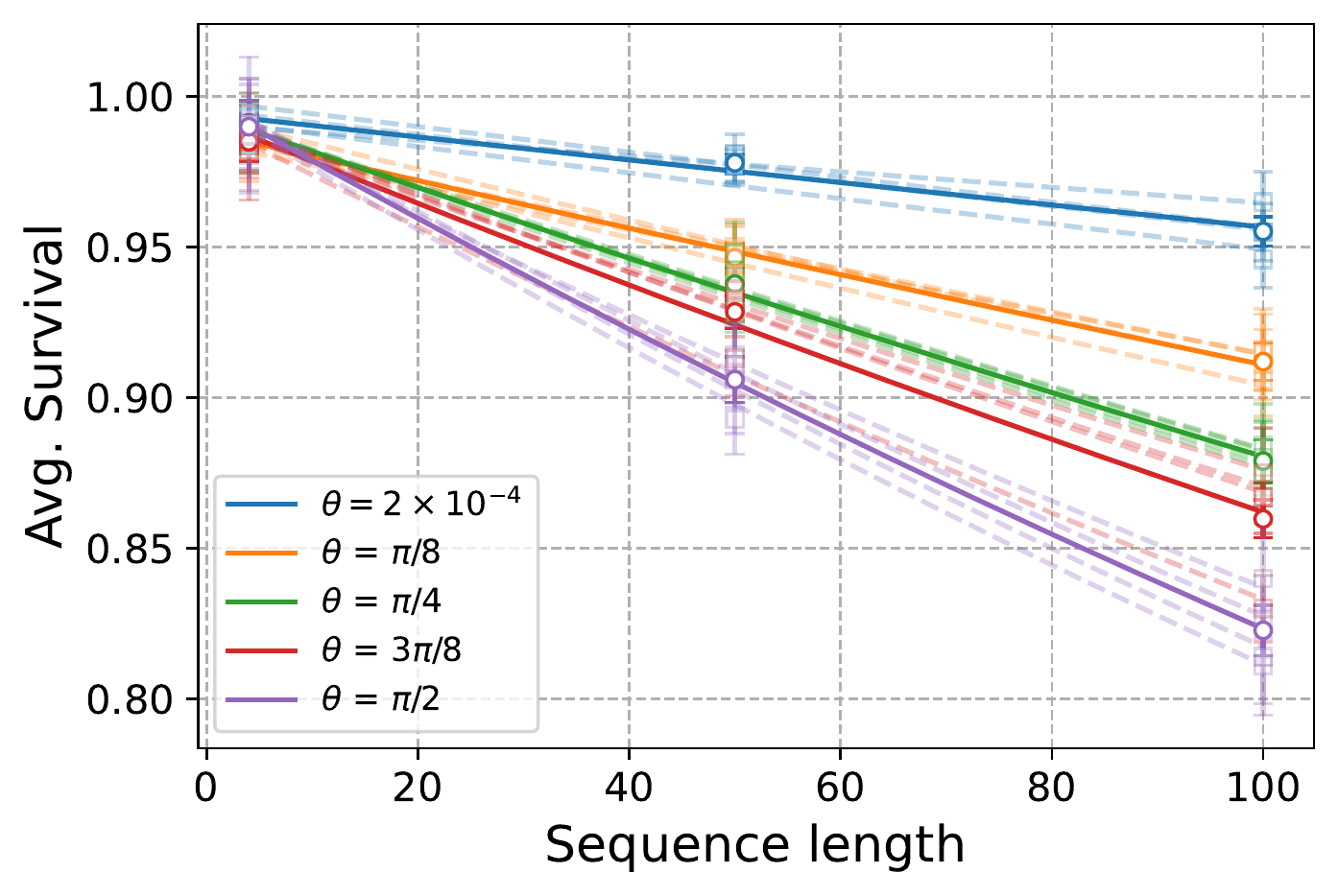}
\caption{Decay curves for direct RB of parameterized 2Q gates.  The different sets of curves show data for $\theta$ ranging from 0 to $\pi/2$ in increments of $\pi/8$. Each experiment used sequence lengths $\ell$=4, 50, and 100, with 10 random circuits per sequence length.  The circuits were run in parallel across the 4 gate zones and all circuits were run in a random order.  The dashed curves are for individual zones, while the solid curve is the average over all zones.  The decay curves for $\theta\in\{\frac{\pi}{8}, \frac{\pi}{4}, \frac{3\pi}{8}, \frac{\pi}{2}\}$ are fit to the model $p(\ell)=Ar^\ell+1/4$, and the average infidelity is computed by Eq.~\eqref{eq: RB average fidelity}. For $\theta=0$, the average infidelity is computed by the procedure described in App.~\ref{app:RB_data}. The average infidelity versus $\theta$ is shown in Fig.~\ref{fig: fid versus angle}.}
\label{fig: arbrb decay}
\end{figure}

Our direct RB circuits are constructed by repeatedly applying $U_{ZZ}(\theta)$ (for a fixed value of $\theta$) interleaved with Haar random SU(2) gates on each qubit. 
For $\theta>0$, this gate set generates SU(4).
The inversion unitary is applied by decomposing the resulting SU(4) element into three $U_{ZZ}(\pi/2)$ 2Q gates using a standard decomposition~\cite{Hanneke2009}.
A final random Pauli is applied to randomize the survival state.
The decay curves are shown in Fig.~\ref{fig: arbrb decay},
and the average fidelity is obtained by fitting to~\eqref{eq: generic RB fit}.

In addition to positive values of $\theta\in\{\frac{\pi}{8}, \frac{\pi}{4}, \frac{3\pi}{8}, \frac{\pi}{2}\}$,
we also run a direct RB experiment with $\theta$ very close to $0$ (specifically $2\times 10^{-4}$),
to measure the baseline error due to
the MS wrapper pulses and memory error accumulated during the cooling pulses.
However, for $\theta=0$, the direct gate set
reduces to SU(2)$\otimes$ SU(2),
which no longer generates a unitary 2-design,
and the RB theory leading to a single exponential decay no longer applies.
To estimate the fidelity in this case,
we use the fact that the action of SU(2)$\otimes$SU(2)
decomposes as a direct sum of 4 irreducible representations (irreps).
(A good introduction to representation theory as it applies to RB is in Ref.~\cite{Helsen2019}.)
The irreps are the span of the identity $II$,
the spans of weight-1 Pauli operators on each qubit $\{IX, IY, IZ\}$ and $\{XI, YI, ZI\}$,
and the span of the weight-2 Pauli operators.
We let $\lambda\in\{II, IZ, ZI, ZZ\}$ label these irreps.
If $\mathcal{E}$ is the error channel for $U_{ZZ}(\theta\approx0)$,
then the twirl of $\mathcal{E}$ over SU(2)$\otimes$ SU(2) is a linear combination of projectors onto these four irreps:
\begin{equation}
    \mathcal{E}_T:=\int_{g\in SU(2)\otimes SU(2)}d\mu(g)\phi(g)\mathcal{E} \phi(g)^{-1} = \sum_{\lambda}r_{\lambda}\Pi_{\lambda},
\end{equation}
where $\phi$ is the superoperator representation of SU(2)$\otimes$ SU(2),
and $\Pi_{\lambda}$ is the projector onto the irrep $\lambda$.
The survival probability at sequence length $\ell$ is then given by
\begin{equation}
p(\ell)=\sum_{\lambda}A_{\lambda}r_{\lambda}^\ell.
\end{equation}
We use the fact that $r_{II}=1$ for trace-preserving maps,
and the randomization in the survival state to fix $A_{II}=1/4$.
To reduce the number of exponential decay curves needed to best-fit to,
we assume qubit symmetry in the error channel,
that is, $r_{IZ}=r_{ZI}=r_1$.
Relabeling the SPAM parameters and defining $r_2=r_{ZZ}$,
the decay model is then given by
\begin{equation}
    p(\ell)=A_1r_1^\ell+A_2r_2^\ell+\frac{1}{4}.
\end{equation}
The entanglement (or process) fidelity is given by
\begin{align}
F&=\frac{1}{16}\sum_{\lambda}\mathrm{dim}(\lambda)r_{\lambda}\notag\\
&= \frac{1}{16}\big(1+6r_1+9r_2\big).
\end{align}
The average infidelity is related to the entanglement fidelity
\begin{equation}\label{eq:ent-to-avg fid}
\epsilon = \tfrac{d}{d + 1}(1 - F),
\end{equation}
for any $d$-dimensional trace-preserving error~\cite{Nielsen2002}.

\section{Details of system-level benchmarks}

\subsection{Mirror benchmarking}

Table~\ref{MB_table} lists the survival probabilities,
decay parameter,
and effective 2Q average infidelity $\epsilon_{\rm eff}^{\rm 2Q}$ for the MB experiment.
The average survival probability as a function
of sequence length $\ell$ is fit to the model
\begin{equation}
    p(\ell)=Au^{\ell-1}.
\end{equation}
Let $\mathcal{E}$ be an $N$-qubit error channel.
Let $\{P_i\}_i$ be the $N$-qubit Pauli operators
with $P_0=\mathds{I}$. The $i$-th Pauli fidelity of $\mathcal{E}$ is defined as
\begin{equation}
    f_i=\frac{1}{2^N}\mathrm{Tr}\big(P_i\mathcal{E}(P_i)\big).
\end{equation}
By applying Pauli randomization to the TQ gates
in the MB circuits,
the error channel for each circuit layer can be assumed to be a stochastic Pauli channel~\cite{Wallman2016}.
Assuming a constant stochastic Pauli error channel $\mathcal{E}$ per circuit layer,
it was shown in Ref.~\cite{Mayer2021} that
the decay parameter $u$ is equal to the
mean square of the non-identity Pauli fidelities:
\begin{equation}
u=\frac{1}{2^N-1}\sum_{i>0}f_i^2.
\end{equation}
For a constant depolarizing error channel on each
2Q gate,
$u$ is given by an analytic formula
(Eq.~(C4) in Ref.~\cite{Mayer2021}).
After best-fitting the experimental decay curves
to obtain $u$,
this formula is used to extract $\epsilon_{\rm eff}^{\rm 2Q}$.

\begin{table}[h!]
\begin{ruledtabular}
\begin{tabular}{ c c c c }
 Sequence length & $N=20$ & $N=26$ & $N=32$ \\ 
 \hline
$\ell=2$ & 0.88(1) & 0.84(2) & 0.82(2) \\
$\ell=4$ & 0.77(3) & 0.70(2) & 0.64(2) \\
$\ell=6$ & 0.66(2) & 0.57(2) & \\
$\ell=7$ & & & 0.51(3) \\
$\ell=10$ & 0.51(3) & 0.39(2) & 0.35(3) \\
\hline
 $u$ & 0.934(6) & 0.908(7) & 0.902(7) \\
 $\epsilon_{\rm eff}^{\rm 2Q}$ & 0.0027(3) & 0.0030(2) & 0.0026(2)
\end{tabular}
\end{ruledtabular}
\caption{\label{MB_table}MB survival probabilities, fit parameter ($u$),
and effective 2Q gate average infidelity ($\epsilon_{\rm eff}^{\rm 2Q}$).
}
\end{table}

\subsection{Quantum volume}
\begin{figure}
\includegraphics[width=0.45\textwidth]{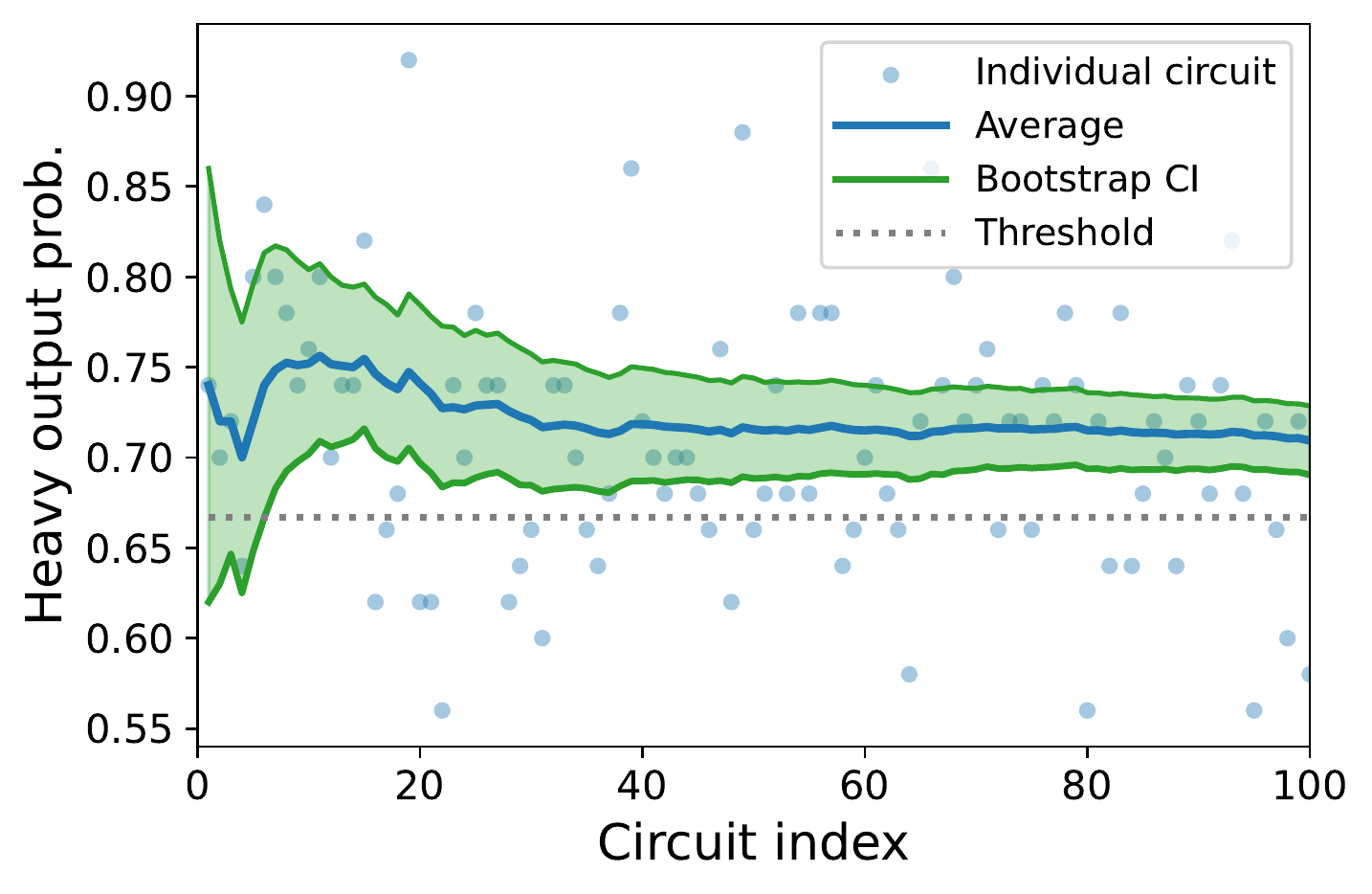}
\caption{Quantum volume $\textrm{QV}=2^{15}$ quantum volume measurement on H2. The average and two-sigma confidence interval of the heavy-output frequency are plotted as a function of the circuit index. Green shaded region shows two-sigma confidence interval from semi-parametric bootstrap method.}
\label{fig:QV15_fig}
\end{figure}

In addition to the $\textrm{QV}=2^{16}$ dataset presented in the main text we also ran several smaller QV tests. In Fig~\ref{fig:QV15_fig} we show the next largest $\textrm{QV}=2^{15}$ test. This test was run with 100 random circuits each with 50 shots and containing an average of 243 parameterized 2Q gates. The measured heavy output probability was 70.9\%, above the threshold with over two-sigma confidence calculated from the semi-parametric bootstrap resampling method.

To infer an effective 2Q error from QV data, first we convert the measured heavy-output probability to a circuit fidelity based on Eq.~13 in Ref.~\cite{Baldwin2022}. We then scale this based on the SPAM error and average number of 2Q gates as shown in Eq.~\ref{eq:crossenttheory}.

\subsection{Random circuit sampling} \label{app:rcs}

The definition of the linear cross-entropy benchmarking fidelity is
\begin{align}
\mathcal{F}_{\text{XEB}} = 2^N \langle P(x_i) \rangle - 1, \label{eq:crossentropy}
\end{align}
where $P(x_i)$ is the probability of measuring the output bitstring $x_i$ in the ideal output distribution, and the expectation value is taken over the empirically measured bitstrings. The linear cross-entropy fidelity is a measure of the correlation between the empirical output distribution and the ideal output distribution. Consequently this requires exact classical simulation of the random circuits, which is a major obstacle to scalability of the benchmark. The uncertainty on the linear cross-entropy fidelity for each circuit can be obtained from \eqref{eq:crossentropy} by combining the variance estimator for $P(x_i)$ with the standard uncertainty-on-the-mean formula, namely,
\begin{align}
    \text{var}( \mathcal{F}_{\text{XEB}} )= \frac{2^{2N} \text{var} (P_i)}{N_{\text{shots}}}.
\end{align}

In Fig.~\ref{fig:crossentropy} we report a fit for the linear cross-entropy benchmarking fidelity on H2 as a function of $N$. This fit was obtained by the following procedure. At each fixed $N$, a representative random circuit was generated and compiled with \verb|pytket| to obtain an expected number of 2Q operations. We note that the final number of 2Q $U_{ZZ}$ operations in each circuit is equal to the number of $\text{fSim} \left(\frac{\pi}{2}, \frac{\pi}{6}\right)$ gates in the original uncompiled circuit. The overall model for the linear cross-entropy fidelity is then
\begin{align} 
\mathcal{F}_{\text{XEB}} &=  (F_{\text{2Q}})^{\# \text{2Q}} \times (1-\epsilon_{\text{SPAM}})^{N}\label{eq:crossenttheory}
\end{align}

Here $F_{\text{2Q}}$ represents the effective entanglement (or process) fidelity of two-qubit operations, while $\epsilon_{\text{SPAM}}$ is the SPAM error as measured by component benchmarking. The conversion between entanglement fidelities and average infidelities as obtained via component benchmarking in Table~\ref{tab:rb_full_data} is given in Eq.~\eqref{eq:ent-to-avg fid}.

\begin{figure*}
\begin{center}
\includegraphics[width=\textwidth]{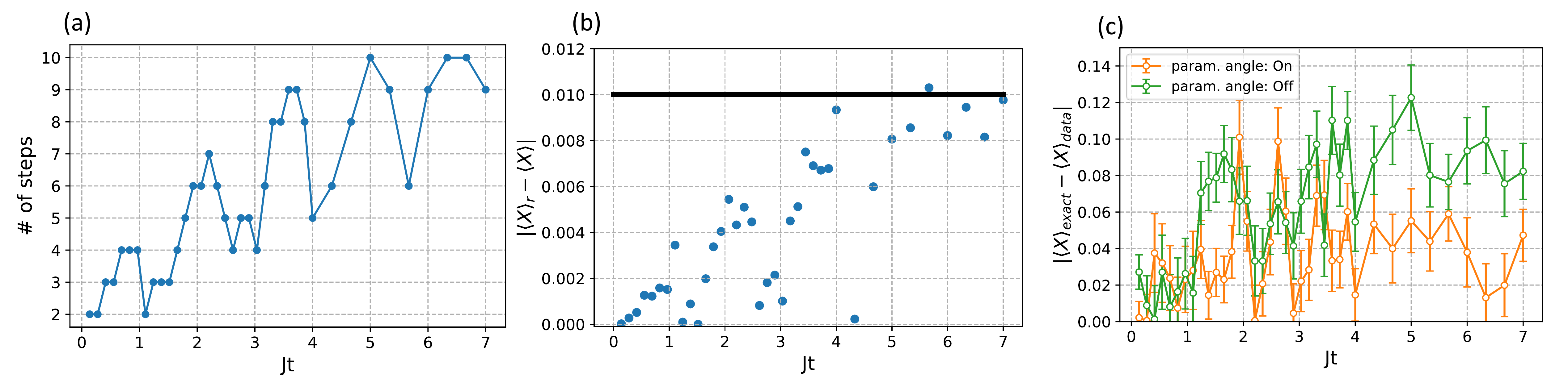} 
\end{center}
\caption{(a) The number of Trotter steps used at each simulation time. (b) The absolute Trotter error $\left|\langle X\rangle_r - \langle X\rangle\right|$ at each time, where $\langle X\rangle_r$ is the expectation value using $r$ Trotter steps under a noiseless circuit and $\langle X \rangle$ is the exact value at that time. (c) The data value relative to the exact value $\left|\langle X\rangle_{\textrm{exact}}-\langle X\rangle_{\textrm{data}}\right|$ at each simulation time. The data error bars are included to reflect signal-to-noise ratio.}
\label{fig:horiz}
\end{figure*}

Holding fixed the average SPAM error of $1.6(1) \times 10^{-3}$ from Table~\ref{tab:rb_full_data}, the model \ref{eq:crossenttheory} was fit to the H2 data, obtaining a best-fit value of $1-F_{\text{2Q}} = 2.4(2) \times 10^{-3}$. In terms of average infidelity, this corresponds to $\epsilon_{\text{2Q}} = 1.9(2) \times 10^{-3}$.

\section{Details of application benchmarks}

\subsection{Trotter steps of Hamiltonian simulation experiment}\label{step_err_detail}

Here we provide details on the Trotter steps used in the experiment. The Trotter steps $r$ are determined by relative convergence with tolerance 0.0025, i.e., we choose a cutoff $r$ such that for $r'\geq r$, neighboring steps are within the threshold $|\<X\>_{r'+1}-\<X\>_{r'}|\leq 0.0025$, where $\<X\>_r$ is the $X$ expectation value after $r$ steps of propagation in a noiseless circuit. For this purpose, we compute each $\<X\>_{r}$ exactly via a discrete-time Jordan-Wigner transformation in the Heisenberg picture \cite{PhysRevA.65.032325}. We checked that this 0.0025 relative error tolerance provides an absolute $\sim 1\%$ Trotter error tolerance in $|\<X\>_{r}-\<X\>|$ for the times we simulate, which is at the scale of the expected $\sim1\%$ statistical fluctuation in the experiment. We chose these values because further improvements from lowering Trotter error would not be reliably observable even if the circuit were completely noiseless, though we did not choose them in a noise-aware fashion (further lowering of the number of Trotter steps used may well give further improvements given the presence of gate errors). The steps and the corresponding Trotter errors are shown in Fig.~\ref{fig:horiz}a,b. The difference between the experiment data and the exact value is shown in Fig.~\ref{fig:horiz}c.

\subsection{The QAOA optimization landscape} \label{app:qaoa}

In Fig.~\ref{fig:N130QAOA} and Fig.~\ref{fig:N32QAOA} in the main text, uncertainties were computed on the expectation value of the energy $\langle H_C \rangle$ as evaluated on H2 (blue points). These uncertainties were computed by bootstrap resampling via the reverse-percentile method \cite{Davison97}, and quantify the uncertainty due to shot noise, but not physical noise sources on the machine. We emphasize that the different data points in Fig.~\ref{fig:N130QAOA} and Fig.~\ref{fig:N32QAOA} are evaluated at different values of the parameters $\vec{\beta}$ and $\vec{\gamma}$.

\begin{figure} 
\includegraphics[width=0.45\textwidth]{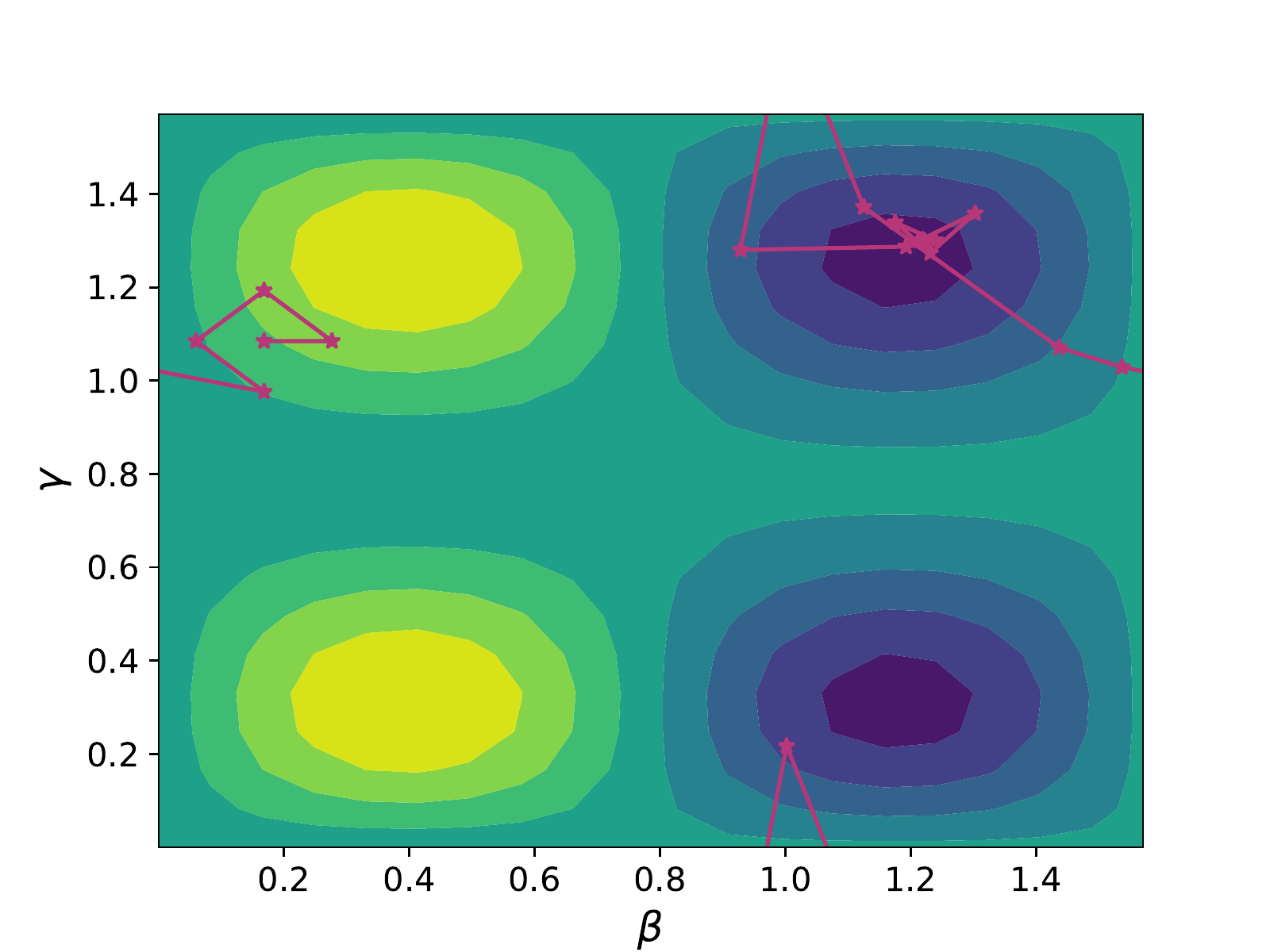}
\caption{The energy landscape of an $N = 130$, $p = 1$  MaxCut QAOA problem. The optimization trace (pink) starts near the maximum in the top left and eventually converges into the potential well in the top right. Individual points at which circuits were evaluated are marked with stars. Note that the landscape is periodic, so the trajectory wraps around the sides of this plot.}
\label{fig:N130Landscape}
\end{figure}

In Fig.~\ref{fig:N130Landscape} we display the full optimization trace on the energy landscape for the $N = 130$, $p = 1$ MaxCut QAOA instance described in the main text, further justifying that the classical optimizer successfully converged to the minimum value of the energy.

\subsection{Details of HoloQUADS experiment}

Here we provide additional details on the holographic quantum dynamics experiments performed on H2. We consider an initial matrix product state of the form
\begin{equation}
	\ket{\psi_0} =\!\!\!\!\!\!\!\! \sum_{\sigma_1,\sigma_2,\dots\in \{\up,\down\}} \!\!\!\!\!\!\!\! \ell^T\mathcal{N}^{(\sigma_1,\sigma_2)}\mathcal{N}^{(\sigma_3,\sigma_4)} \cdots  \ket{\sigma_1\sigma_2\sigma_3\sigma_4 \cdots }.\label{eq:solvableMPS}
\end{equation}
with the tensor $\N^{(\sigma,\sigma')}_{i,j} = \bra{j} \otimes \bra{\sigma'} W \ket{i} \otimes \ket{\sigma}$ specified by the unitary  $W=\exp[-i(K_x XX + K_y YY + K_z ZZ)]$ with $(K_x,K_y,K_z)=(0.3,0.5,1.25)$; this bond-dimension $\chi=2$ matrix product state, previously studied in Refs.~\cite{Bertini2019,HoloChertkov2022}, is prepared by applying gates between the physical qubits and $n_b = \log_2 \chi=1$ ancilla ``bond'' qubits. We time-evolve this state by the SDKI model \cite{Akila2016,bertini2018}, which can be formulated as a dual-unitary circuit using 2Q gates
\begin{equation}
U = (u_+ \otimes u_-) e^{-i\frac{\pi}{4}\left(XX + YY \right)}(v_- \otimes v_+) .\label{eq:du_gate}
\end{equation}
Here the 1Q gates are given by $u_+ = e^{-ihZ}e^{i\frac{\pi}{4}X}e^{-i\frac{\pi}{4}Y}$, $u_- = e^{i\frac{\pi}{4}X}e^{-i\frac{\pi}{4}Y}, v_- = e^{i\frac{\pi}{4}Y}e^{-ihZ}$, and $v_+=e^{i\frac{\pi}{4}Y}$. We used $h=0.05$, which is close to but not exactly at the integrable $h=0$ point where the SDKI model displays no decay of correlation functions.

Qubit-reuse techniques involving MCMR~\cite{HoloChertkov2022,DeCross2022} are used to construct the circuit so the four gate zones in H2 are used as in-parallel as possible. The leakage detection gadget (Fig.~\ref{leak_det_fig}) is used to discard results where the bond qubit was measured to have leaked (measuring $2\%,2\%,5\%,7\%$ bond qubit leakage for $t=0,8,16,24$). We also use a circuit identity to construct each gate $U$ with a single parameterized $U_{ZZ}(\theta)$ gate and one physical SWAP~\cite{HoloChertkov2022}.

\end{document}